\documentclass[twocolumn,preprintnumbers,amsmath,amssymb,superscriptaddress,nofootinbib,english]{revtex4-1}
\pdfoutput=1
\usepackage{times,amsmath,amsfonts,amssymb,epstopdf}
\usepackage{graphicx}
\usepackage{dcolumn}
\usepackage{bm}
\usepackage{enumerate}
\usepackage{epsfig}
\usepackage{graphicx}
\usepackage{hyperref}
\usepackage[usenames]{color}
\usepackage{url}
\usepackage[normalem]{ulem}
\usepackage[T1]{fontenc}
\usepackage{amsmath}

\usepackage[usenames]{color}

\def\be{\begin{equation}}
\def\ee{\end{equation}}
\def\ben{\begin{eqnarray}}
\def\een{\end{eqnarray}}
\def\ba{\begin{array}}
\def\ea{\end{array}}

\def\ba#1\ea{\begin{align}#1\end{align}}

\newcommand{\bq}{\begin{eqnarray}}
\newcommand{\eq}{\end{eqnarray}}
\newcommand{\bes}{\begin{subequations}}
\newcommand{\ees}{\end{subequations}}

\begin{document}
\newcommand{\half}{{\textstyle\frac{1}{2}}}
\allowdisplaybreaks[3]
\def\triangledown{\nabla}
\def\grad3{\hat{\nabla}}
\def\a{\alpha}
\def\b{\beta}
\def\g{\gamma}\def\G{\Gamma}
\def\d{\delta}\def\D{\Delta}
\def\ep{\epsilon}
\def\et{\eta}
\def\z{\zeta}
\def\t{\theta}\def\T{\Theta}
\def\l{\lambda}\def\L{\Lambda}
\def\m{\mu}
\def\f{\phi}\def\F{\Phi}
\def\n{\nu}
\def\r{\rho}
\def\s{\sigma}\def\S{\Sigma}
\def\ta{\tau}
\def\x{\chi}
\def\o{\omega}\def\O{\Omega}
\def\k{\kappa}
\def\pa {\partial}
\def\ov{\over}
\def\br{\\}
\def\ud{\underline}

\def\lcdm{\Lambda{\rm CDM}}
\def\msun{M_{\odot}/h}
\def\sdgp{\rm DGPs}
\def\mdgp{\rm DGPm}
\def\wdgp{\rm DGPw}

\def\mnras{MNRAS}

\newcommand{\comment}[1]{}

\title{Validating estimates of the growth rate of structure with modified gravity simulations}

\author{Alexandre Barreira}
\email[Electronic address: ]{barreira@MPA-Garching.MPG.DE}
\affiliation{Max-Planck-Institut f{\"u}r Astrophysik, Karl-Schwarzschild-Str. 1, 85741 Garching, Germany}

\author{Ariel G. S\'{a}nchez}
\email[Electronic address: ]{arielsan@mpe.mpg.de}
\affiliation{Max-Planck-Institut f{\"u}r extraterrestrische Physik, Giessenbachstr., 85741 Garching, Germany}

\author{Fabian Schmidt}
\email[Electronic address: ]{fabians@MPA-Garching.MPG.DE}
\affiliation{Max-Planck-Institut f{\"u}r Astrophysik, Karl-Schwarzschild-Str. 1, 85741 Garching, Germany}

\begin{abstract}

We perform a validation of estimates of the growth rate of structure, described by the parameter combination $f\sigma_8$, in modified gravity cosmologies. We consider an analysis pipeline based on the redshift-space distortion modelling of the clustering wedges statistic of the galaxy correlation function and apply it to mock catalogues of $\lcdm$ and the normal branch of DGP cosmologies. We employ a halo occupation distribution approach to construct our mocks, which we ensure resemble the CMASS sample from BOSS in terms of the total galaxy number density and large scale amplitude of the power spectrum monopole. We show that the clustering wedges model successfully recovers the true growth rate difference between DGP and $\lcdm$, even for cases with over 40\% enhancement in $f\sigma_8$ compared to $\lcdm$. The unbiased performance of the clustering wedges model allows us to use the growth rate values estimated {from the BOSS DR12 data to constrain the cross-over scale $r_c$ of DGP gravity to $\left[r_cH_0\right]^{-1} < 0.97$ ($2\sigma$) or $r_c > 3090\ {\rm Mpc}/h$, cutting into the interesting region of parameter space with $r_c \sim H_0^{-1}$ using constraints from the growth of structure alone.}
\end{abstract} 

\maketitle

\section{Introduction}\label{sec:intro}

Cosmological studies of theories of gravity beyond General Relativity (GR) have become the focus of growing interest in the past few years (see Refs.~\cite{2012PhR...513....1C, Joyce:2014kja, 2016arXiv160106133J} for recent reviews). One of the main reasons for this is that these {\it modified gravity} scenarios may offer an explanation for the observed accelerated expansion of the Universe that does not invoke the existence of exotic dark energy components or a finely-tuned cosmological constant $\Lambda$. Another main source of interest is related to the usefulness of these models in the design of cosmological tests of gravity: the study of the phenomenology of alternative models helps to determine the types of observational signatures that one should be looking for in the data gathered by current and ongoing observational missions such as DES \cite{Abbott:2015swa}, BOSS \cite{2014MNRAS.443.1065B}, Euclid \cite{2011arXiv1110.3193L}, LSST \cite{2012arXiv1211.0310L} and DESI \cite{2013arXiv1308.0847L}. For example, a general prediction of modified gravity theories is the existence of a long-range {\it fifth force} that universally couples to matter \cite{2016arXiv160106133J}.  This force has an impact on the peculiar motion of galaxies \cite{2014PhRvL.112v1102H} and consequently leaves an inprint on the anisotropic galaxy clustering pattern induced by redshift-space distortions (RSD).

{The statistics of galaxy clustering are isotropic in a freely falling Friedmann-Robertson-Walker (FRW) frame with comoving coordinates.} The measured redshifts of galaxies, however, are affected by their peculiar velocities, which causes an anisotropy in the observed galaxy statistics that is proportional to the
velocity of galaxies \cite{Kaiser:87}.  If galaxy velocities are unbiased with respect to matter, which is expected to be the case on large scales {\cite{MSZ,senatore:2015}}, then this allows for a measurement
of the growth rate of structure $f = {\rm dln}D/{\rm dln}a$, where $D$ is the growing mode of linear density fluctuations and $a$ the cosmological scale factor
(see Ref.~\cite{1998ASSL..231..185H} for a review).  Recent such analyses include those of the 6dF survey \cite{2012MNRAS.423.3430B}, the luminous red galaxy (LRG) sample of the DR7 from the Sloan Digital Sky Survey (SDSS) \cite{2014MNRAS.439.2515O}, the galaxy samples from the BOSS survey \cite{2013arXiv1312.4889C, 2014MNRAS.439.3504S, 2013MNRAS.433.1202S, 2014MNRAS.440.2692S, 2014MNRAS.443.1065B, 2015arXiv150906386G, 2016arXiv160703147S}, the VIPERS survey \cite{2013A&A...557A..54D} and the WiggleZ survey \cite{2012MNRAS.425..405B}.  A set of constraints on $f(z_i)$ at different redshifts $z_i$ can then in principle be used to constrain various modified gravity models, whose predictions for $f$ can be calculated using linear theory.

As statistical uncertainties become smaller, it becomes increasingly pressing to determine the importance of model systematics in our scrutiny of cosmological data. Hence, before applying any given observational analysis pipeline, one should first validate it against N-body simulations and check whether it is able to recover the growth rate of the input cosmology in an unbiased way. Moreover, these checks should be performed for as many classes of cosmological scenarios one wishes to test in order to ensure a fair comparison between theory and observations. 

So far, most of the existing RSD models have been validated only against simulations of cosmologies assuming GR, but a few steps in the direction of understanding and developing RSD models for modified gravity have already been taken. For instance, Ref.~\cite{2012MNRAS.425.2128J} studied RSD using N-body simulations of the Hu-Sawicki $f(R)$ theory of gravity \cite{Hu:2007nk} and $\lcdm$. The authors found that the same RSD model for the two-dimensional redshift power spectrum does not exhibit exactly the same level of agreement with the simulations of the two cosmologies. In Ref.~\cite{2014PhRvD..89d3509T}, the authors have generalized RSD models based on perturbation theory to include the effects of modified gravity. With the aid of N-body simulations of $f(R)$, the authors have further shown that the improved model does help to reduce the bias in the recovery of the input $f(R)$ cosmology. This improved model has been applied to real data to place constraints on the $f(R)$ model in Ref.~\cite{2015PhRvD..92d3522S} (see also Ref.~\cite{hojjati/etal}). Reference \cite{2013PhRvD..88h4029W} has also studied redshift-space distortions in a Galileon-like theory of gravity.  Constraints on the Dvali-Gabadadze-Porrati (DGP) \cite{Dvali:2000hr} model studied here have been placed
using RSD on linear scales in Ref.~\cite{2013MNRAS.436...89R}.  

In this paper, we work with RSD models of the clustering wedges statistic of the two-point galaxy correlation function. {In Ref.~\cite{2016arXiv160703147S}, the analysis pipeline that we use in this paper has been shown to return unbiased constraints on the cosmological parameters of a suite of $\lcdm$ mock catalogues (see also Refs.~\cite{2013MNRAS.433.1202S, 2014MNRAS.440.2692S})}. Here, we wish to determine to which extent this model of the clustering wedges is able to also successfully recover the growth rate in modified gravity cosmologies. We take as our working case cosmology the normal branch of DGP braneworld gravity and apply the RSD model to CMASS-like mock catalogues, which we build out of N-body simulations of modified gravity \cite{codecomp}.  Such a test at the level of full modified gravity mock catalogs has not been carried out previously.  We shall find that the clustering wedges model (which has not been augmented with any ingredient to specifically account for modified gravity) shows no evidence of retrieving biased estimates of $f$, even when applied to quite dramatic fifth force effects. This result indicates that the growth rate estimated {from the BOSS DR12 galaxy sample with this pipeline \cite{2016arXiv160703147S} can be used to place robust constraints on DGP models, and on models with similar phenomenology.}

The rest of this paper is organized as follows. In Sec.~\ref{sec:dgp}, we summarize the main aspects of DGP gravity cosmologies. In Sec.~\ref{sec:sims}, we describe our N-body simulation setup and explain how we construct our galaxy mock catalogues. In Sec.~\ref{sec:main}, we describe the clustering wedges model used in our analysis and present the constraints on the growth rate from our mock catalogues. We also use the growth rate estimates obtained from the real data to place constraints on the DGP model. We summarize and conclude in Sec.~\ref{sec:conc}. 

If not specified, we work in units in which $c=1$.

\section{Working case cosmology: DGP gravity}\label{sec:dgp}

We consider DGP braneworld gravity scenarios \cite{Dvali:2000hr}, which are one of the most thoroughly studied modified gravity cosmologies in terms of structure formation in both the linear and nonlinear regimes \cite{2009PhRvD..80d3001S, 2009PhRvD..80l3003S, baojiudgp, 2012PhRvL.109e1301L, 2013MNRAS.436...89R, 2014MNRAS.445.1885Z, 2014JCAP...07..058F, 2015JCAP...07..049F, codecomp, 2015JCAP...12..059B}. Specifically, we shall focus on the stable normal branch of the theory and include a dark energy component adjusted to yield an expansion history identical to $\lcdm$ \cite{2009PhRvD..80l3003S,SahniShtanov}. This model introduces one additional parameter over $\lcdm$, which is called the cross-over scale $r_c$. In the limit $r_c \to\infty$ one recovers $\lcdm$.

While theoretically not very appealing, this model satisfies constraints from the expansion history as well as Solar System tests, making it a very useful toy model to constrain using large-scale structure. Next, we briefly introduce this model and display the relevant equations.

\subsection{Action and background evolution}\label{sec:bck}

The action of the DGP model is split into a four- and a five-dimensional part
\bq\label{eq:dgpaction}
S &=& \int_{\rm brane}\!\!\! {\rm d}^4x \sqrt{-g} \left(\frac{R}{16\pi G} + \mathcal{L}_m\right) \nonumber \\
&+& \int {\rm d}^5x \sqrt{-g^{(5)}} \left(\frac{R^{(5)}}{16\pi G^{(5)}}\right),
\eq
where $g^{(5)}$ and $g$ are, respectively, the determinants of the metric of the five-dimensional bulk $g_{\mu\nu}^{(5)}$ and four-dimensional {\it brane} $g_{\mu\nu}$, and $R^{(5)}$ and $R$ are their corresponding Ricci scalars. $\mathcal{L}_m$ represents the Lagrange density of any energy component (matter, radiation, dark energy) that exists on the brane. The two gravitational strengths $G^{(5)}$ and $G$ can be used to define the parameter $r_c$, 
\bq\label{eq:rc}
r_c = \frac{1}{2}\frac{G^{(5)}}{G}.
\eq

The background expansion rate in the DGP model can be written as\footnote{Neglecting the contribution from radiation, which is negligible at the cosmological times we shall be interested in.} \cite{2001PhLB..502..199D, DeffayetEtal02, 2007PhRvD..75h4040K, 2006JCAP...01..016K} 
\bq\label{eq:dgpH}
H(a) = H_0\sqrt{\Omega_{m0}a^{-3} + \Omega_{\rm de}(a) + \Omega_{rc}} \pm \sqrt{\Omega_{rc}},
\eq
where $H_0 = 100h\ {\rm km/s/Mpc}$ is the expansion rate today, $\Omega_{m0} = 8\pi G \bar{\rho}_{m0}/(3H_0^2)$ is the fractional matter density today, $\bar{\rho}_{m0}$ is the present-day value of the background matter density $\bar{\rho}_{m}$, $\Omega_{rc} = 1/(4H_0^2r_c^2)$ and $\Omega_{\rm de}(a)$ is the fractional energy density of some dynamical dark energy field that may exist on the brane. The choice of the sign of the second term on the right-hand side of Eq.~(\ref{eq:dgpH}) defines two branches of the model. In the {\it self-accelerating branch} ($-$ sign), the expansion of the Universe accelerates at late times even if there is no explicit dark energy component on the brane, $\Omega_{\rm de} = 0$ (hence the name). This case is however ruled out by cosmic microwave background (CMB) and supernovae data \citep{2008PhRvD..78j3509F}, and it suffers also from a number of theoretical instabilities \citep{2003JHEP...09..029L,2004JHEP...06..059N, 2007CQGra..24R.231K}. The so-called {\it normal branch} ($+$ sign) does not suffer from such observational and theoretical problems, but requires $\Omega_{\rm de}(a) \neq 0$ to drive the accelerated expansion.  From hereon in this paper, we consider this stable normal branch and assume a tuned time evolution of $\Omega_{\rm de}(a)$ such that overall $H(a) = H_{\lcdm}(a)$ \cite{2009PhRvD..80l3003S}.  As shown in Ref.~\cite{2009PhRvD..80l3003S}, the equation of state of this dark energy component is always greater than $-1$, and it reduces to a cosmological constant in the limit $r_c \to \infty$.  By tuning $\Omega_{\rm de}(a)$ in such a way, the changes to structure formation w.r.t.~$\lcdm$ come solely from the fifth force (described below) and not from the modified background dynamics.

\subsection{Structure formation}\label{sec:sf}
In the normal branch of DGP gravity, structure formation on scales much smaller than both the horizon and the cross-over scale is governed by the equations
\bq
\label{eq:modpoisson}&&\nabla^2\Psi = 4\pi G a^2 \delta\rho_m + \frac{1}{2}\nabla^2\varphi, \\
\label{eq:eomvarphi}&&\nabla^2\varphi + \frac{r_c^2}{3\beta(a) a^2} \left[\left(\nabla^2\varphi\right)^2 - \left(\nabla_i\nabla_j\varphi\right)^2\right] = \frac{8\pi G}{3\beta(a)}a^2 \delta\rho_m, \nonumber \\
\eq
where $\varphi$ is a scalar degree of freedom associated with the bending modes of the brane and $\delta\rho_m = \rho_m - \bar{\rho}_{m}$ is the matter density perturbation (an overbar indicates background averaged quantities). These equations correspond to a perturbed FRW metric on the brane (considering only scalar perturbations {and assuming spatial flatness})
\bq\label{eq:metric}
{\rm d}s^2 = \left(1 + 2\Psi\right){\rm d}t^2 - a(t)^2\left(1 - 2\Phi\right){\rm d}x^2,
\eq
and are derived under the so-called (i) quasi-static approximation, which amounts to neglecting time-derivatives of $\varphi$ over spatial ones; and (ii) weak-field limit, $\Psi, \Phi, \varphi \ll 1$ (assuming the same boundary conditions for the gravitational potentials and the scalar field). References \cite{2009PhRvD..80d3001S, 2014PhRvD..90l4035B, 2015arXiv150503539W} have verified the validity of these two approximations, which are standard in modified gravity studies (see also Ref.~\cite{2014PhRvD..89b3521N}). We have also neglected perturbations in the dark energy component, which are negligible on the sub-horizon scales we are interested in.  In Eq.~(\ref{eq:eomvarphi}), $\beta(a)$ is given by
\bq\label{eq:beta}
\beta(a) = 1 + 2Hr_c\left(1 + \frac{\dot{H}}{3H^2}\right),
\eq
where the dot denotes a derivative w.r.t.~physical time $t$. In DGP gravity, only the dynamical potential gets modified w.r.t.~GR, $\Psi = \Psi^{\rm GR} + \varphi/2$, with the lensing potential remaining the same, $\Phi_{\rm len} = \left(\Phi + \Psi\right)/2 = \Phi^{\rm GR}$.

The growing mode, $D$, of linear density fluctuations on sub-horizon scales in DGP cosmologies is governed by
\bq\label{eq:linearD}
\ddot{D} + 2H\dot{D} - 4\pi G_{\rm eff}^{\rm lin}\bar{\rho}_m D = 0,
\eq
where $G_{\rm eff}^{\rm lin} = G\left[1 + 1/(3\beta)\right]$ is the linear effective gravitational strength (defined as $\nabla^2\Psi = 4\pi G_{\rm eff}\delta\rho_m$, after combining the linearized form of Eqs.~(\ref{eq:modpoisson}) and (\ref{eq:eomvarphi})). Note that $G_{\rm eff}^{\rm lin}$ is a function of time only, which means that the linear growth of structure is scale-independent, as in GR.

\subsection{Screening mechanism}\label{sec:screening}

The nonlinear derivative terms in Eq.~(\ref{eq:eomvarphi}) give rise to a screening effect known as the Vainshtein mechanism \cite{Vainshtein1972393, Babichev:2013usa, Koyama:2013paa}, which is what gives the DGP model a chance to pass the stringent Solar System tests of gravity \cite{Will:2014xja}.  In order to build an intuition for how the Vainshtein screening works, it is best to assume spherical symmetry. In this case, Eq.~(\ref{eq:eomvarphi}) can be immediately integrated once over the physical radial coordinate $r$ to become:
\bq\label{eq:eomvarphi_sph2}
\varphi,_r = \frac{4}{3\beta}\left(\frac{r}{r_V}\right)^3\left[-1 + \sqrt{1 + \left(\frac{r_V}{r}\right)^3}\right]\frac{GM(r)}{r^2},
\eq
where
\bq\label{eq:rv}
r_V(r) = \left(\frac{16r_c^2GM(r)}{9\beta^2}\right)^{1/3}
\eq
is a distance scale known as the {\it Vainshtein radius}, $M(<r) = 4\pi\int_0^r r'^2\rho(r'){\rm d}r'$ and a comma denotes partial differentiation w.r.t.~$r$. The value of $r_V(r)$ determines the distance scale to the source below which the effects of the fifth force $F_{5 \rm th} = \varphi,_r/2$ become suppressed relative to the standard GR contribution. Consider for simplicity a top-hat profile with size $R_{\rm TH}$ and mass $M_{\rm TH}$. If $r \gg r_V(r) > R_{\rm TH}$, then
\bq\label{eq:forcelarge}
F_{5\rm th} = \frac{\varphi,_r}{2} \approx \frac{1}{3\beta}\frac{G M_{\rm TH}}{r^2} = \frac{1}{3\beta}F_{\rm GR}, 
\eq
i.e. the contribution of $\varphi$ is comparable to that of standard GR ($\beta \sim \mathcal{O}(1)$ at late times for $r_c H_0\sim 1$). On the other hand, for  $R_{\rm TH} < r \ll r_V(r)$, we have
\bq\label{eq:forcesmall}
\frac{F_{5\rm th}}{F_{GR}} \rightarrow 0, \ \ \ \ \ \ {\rm as} \ \ \ \ \ \frac{r}{r_V} \rightarrow 0,
\eq
i.e., the fifth force has a negligible impact, thereby permiting the model to meet all Solar System bounds.

\section{HOD galaxy catalogues}\label{sec:sims}

In this section, we describe our simulations and the steps performed to construct the galaxy mock catalogues.

\subsection{N-body simulations of DGP gravity}\label{sec:simsdgp}

The N-body simulations we use in this paper were run with the {\tt ECOSMOG} code \cite{2012JCAP...01..051L}, which is a modified version of the publicly-available Adaptive Mesh Refinement (AMR) {\tt RAMSES} code \cite{2002A&A...385..337T}. More specifically, we use the version of {\tt ECOSMOG} developed for simulations of models with Vainshtein screening \cite{baojiudgp, 2013JCAP...11..012L} and employ the speed-up method described in Ref.~\cite{2015JCAP...12..059B}.  This code was cross-checked with other implementations of the DGP equations of motion in Ref.~\cite{codecomp}, finding good agreement.  In short, the code discretizes  Eq.~(\ref{eq:eomvarphi}) on the AMR grid and solves it using Gauss-Seidel iterations. This yields the scalar field value at every cell of the AMR grid, which is used to compute the fifth force ($\vec{\nabla}\varphi$) via finite-differencing. We used a Cloud-in-Cell (CIC) scheme to interpolate the forces from the cell centers (where they are evaluated) to the particle positions. To ensure momentum conservation, we construct the density field on the AMR grid (which sources the standard and fifth forces) also with a CIC interpolation scheme from the particle positions. The speed-up method is implemented by truncating the Gauss-Seidel iterations above a certain AMR refinement level. On these higher refinements, the local density is high enough for the screening mechanism to be very effective. Hence, given that the relative contribution of the fifth force is small, it is a good approximation to interpolate the scalar field values on higher levels from its solutions on coarser levels. The error this induces on the total force (GR plus fifth force) is small, but the improvement in the performance of the code is very significant (see Ref.~\cite{2015JCAP...12..059B} for more details). This therefore enables the use of larger box sizes and with better mass resolution in modified gravity simulations.

The initial conditions were generated at $z = 49$ using the following cosmological parameter values (from the first column in Table 3 of the Planck mission constraints paper \cite{2015arXiv150201589P})
\bq\label{eq:cosmoparams}
&&\left\{\Omega_{b0}, \Omega_{c0}, h, n_s, A_s10^{9} \right\} = \\ \nonumber
&&\left\{0.049, 0.2642 , 0.6731, 0.9655, 2.195\right\},
\eq
where $\Omega_{b0}, \Omega_{c0}, h, n_s, A_s$ are, respectively, the fractional baryon density today, the fractional dark matter density today, the dimensionless Hubble expansion rate (already introduced above), the primordial spectral index and the amplitude of the primordial matter power spectrum at a pivot scale $k = 0.05\ {\rm Mpc}^{-1}$. {The Planck analysis of Ref.~\cite{2015arXiv150201589P} places bounds on these parameters assuming that the Universe is $\lcdm$. Since the $\lcdm$ and DGP models studied here are indistinguishable at early times and share the same expansion history, we expect that the best-fitting values for $\lcdm$ lead also to a reasonable fit to the CMB data in DGP cosmologies (even if the corresponding bounds may become looser because of the impact of the fifth force at late times on the integrated Sachs-Wolfe (ISW) effect and also on CMB lensing). For simplicity, we therefore use the same set of parameter values to simulate both $\lcdm$ and DGP. We note, anyway, that for our validation analysis it is not critical to require that the simulated cosmologies are very good fits to the CMB.} We consider three values of the cross-over scale: $r_cH_0 = 0.1$, $r_cH_0 = 0.5$ and $r_cH_0 = 2.0$, which we call, respectively, DGPs(trong), DGPm(edium), and DGPw(eak), referring to the strength of the fifth force. We note that although models like $\sdgp$ or $\mdgp$ may already be too extreme to be compatible with current data (see below and Ref.~\cite{2013MNRAS.436...89R}), we include them in our analysis anyway {to increase the range of fifth force strengths explored in our validation tests.} For comparison purposes, we also simulate  a $\lcdm$ cosmology. Some specifications of the models we consider in this paper are summarized in Table \ref{table:models}.

We employ a simulation box with side $L_{\rm box} = 600{\rm Mpc}/h$ and $N_p = 1024^3$ dark matter tracer particles. The cell size of the {first refined AMR level is $l_{\rm first} = L_{\rm box} / N_p^{1/3} / 2 \approx 0.3 {\rm Mpc/h}$ which is small enough to capture the suppression of fifth forces due to the screening mechanism \cite{2015JCAP...12..059B}. For this reason and to improve the performance of the code, we only explicitly solve the fifth force on the domain (that which regularly covers the whole box) and first refined levels, and interpolate this solution to all other finer levels \footnote{{We have explicitly checked that mock catalogues constructed from simulations where the fifth force is only solved on the domain level are indistinguishable from those presented in this paper (cf.~Sec.~\ref{sec:hodmodel}). This ensures that our mocks are not affected by numerical artifacts coming from the speed-up method.}}}. Note that a modified gravity simulation of this resolution would be extremely computationally demanding if the scalar field iterations would take place on all AMR levels. In our simulations, the AMR cells get refined (de-refined) if the effective number of particles contained in its spatial volume is larger (smaller) than $8$. All our simulations evolve from the exact same set of initial conditions, a fact that we shall take into consideration below when determining the statistical error of our measurements.

\subsection{Halo mass function and linear bias}\label{sec:mf}

\begin{figure}
	\centering
	\includegraphics[scale=0.49]{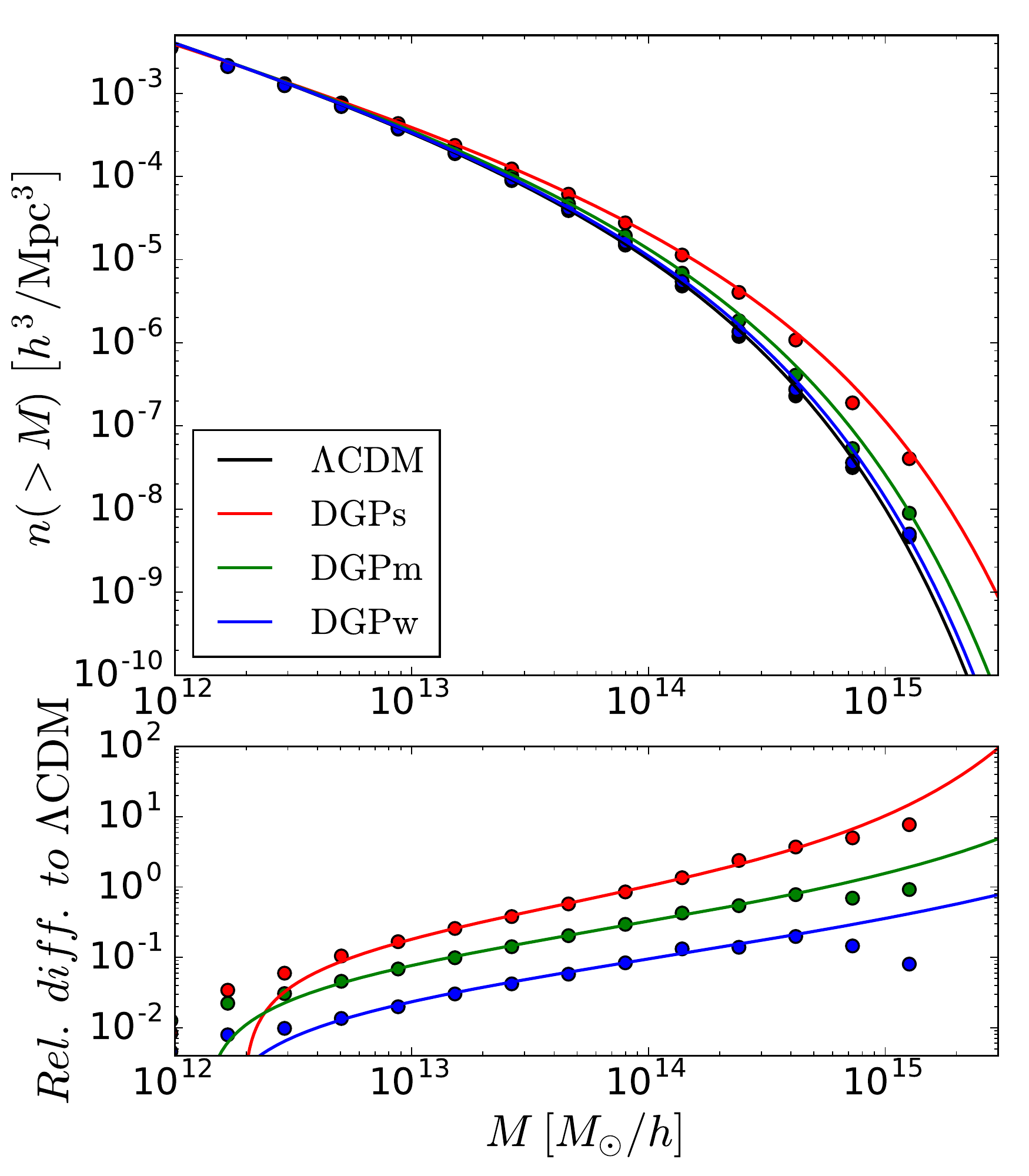}
	\caption{Cumulative halo mass function of $\lcdm$ and the three DGP cases at $z = 0.57$, as labelled (the upper panel shows the absolute value, while the lower panel shows the relative difference to $\lcdm$). The dots show the simulation results and the solid lines display the best-fitting ST formulae. These halo catalogues were constructed with the {\tt Rockstar} code \cite{2013ApJ...762..109B}.}
\label{fig:cmf}
\end{figure}

\begin{figure}
	\centering
	\includegraphics[scale=0.40]{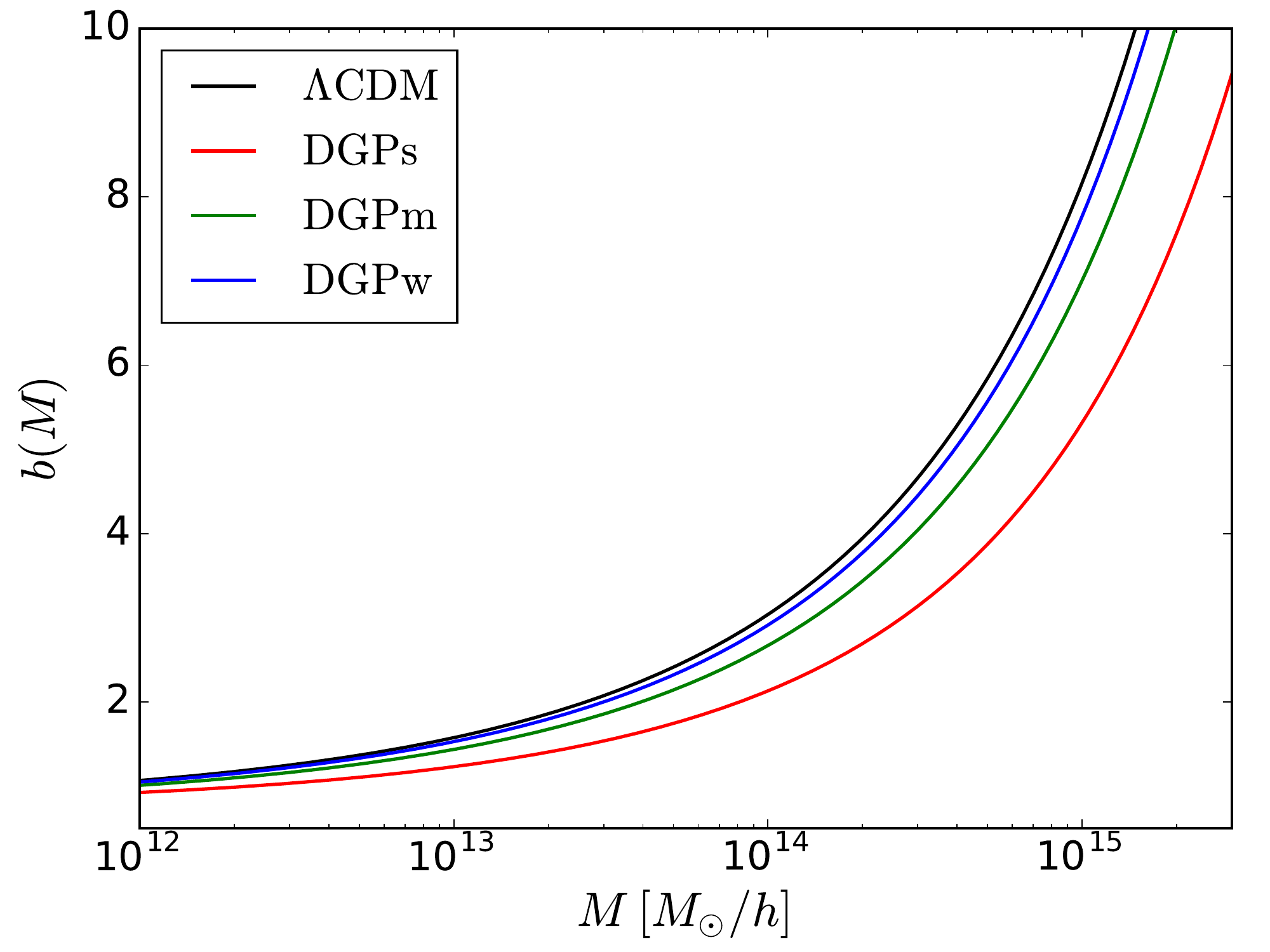}
	\caption{Best-fitting ST linear halo bias of $\lcdm$ and the three DGP cases at $z = 0.57$, as labelled.}
\label{fig:hb}
\end{figure}

Figure \ref{fig:cmf} shows the cumulative halo mass function at $z = 0.57$ (the mean redshift of the CMASS sample) measured from the $\lcdm$, $\sdgp$, $\mdgp$ and $\wdgp$ simulations (dots), as labelled. The solid lines correspond to the best-fitting Sheth-Tormen (ST) mass function formulae (cf.~Eqs.~(\ref{eq:mass-function}) and (\ref{eq:first-crossing-ST})). We refer the interested reader to Appendix \ref{sec:stfit} for the fitting procedure of the ST formulae to the simulation results.  Note that the worse performance of the ST mass function in fitting the deviation from $\lcdm$ for $M \lesssim 5 \times 10^{12} M_\odot/h$ is not important in practice, as the difference in the mass function is at the few percent level only. The ST linear halo bias is shown in Fig.~\ref{fig:hb}.

These two figures show the expected result that the effects of the positive fifth force boost the number of massive halos and reduce their linear bias values \cite{2009PhRvD..80l3003S, 2013PhRvD..88h4029W, 2014JCAP...04..029B}, relative to $\lcdm$. This has an impact on the way galaxies populate dark matter halos in order to match the observed clustering properties of a given galaxy sample, as we shall see next.

\subsection{Halo Ocupation Distribution model}\label{sec:hodmodel}

\begin{figure*}
	\centering
	\includegraphics[scale=0.415]{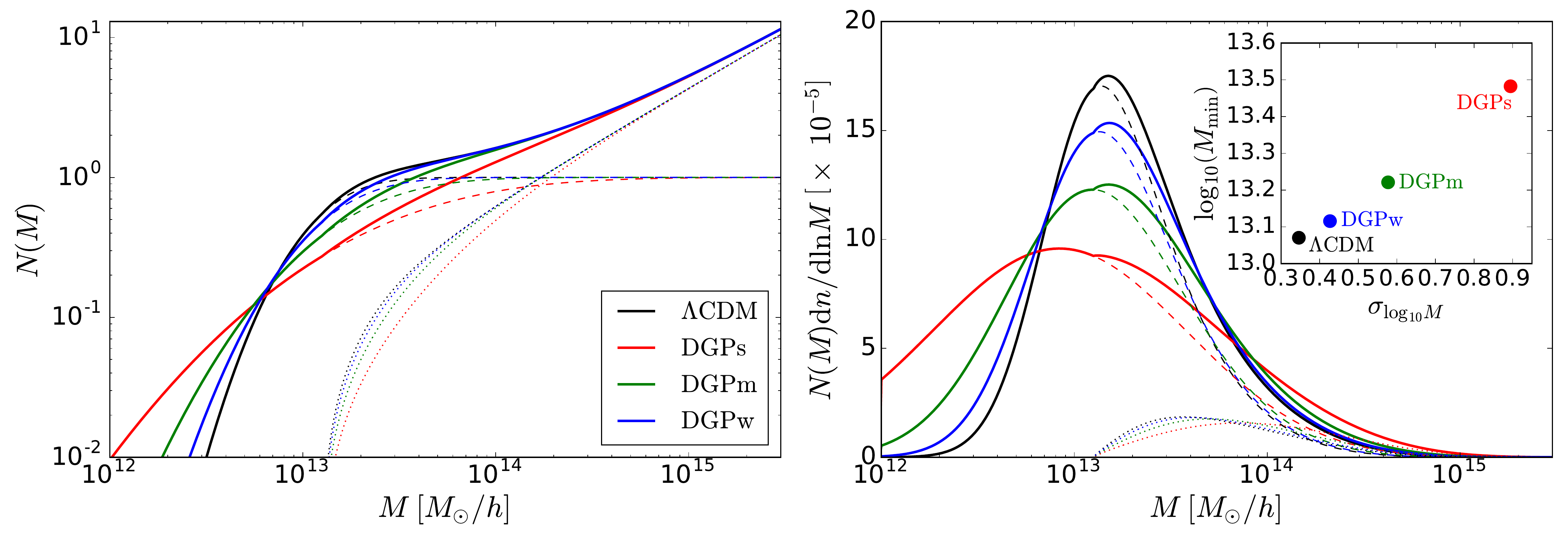}
	\caption{Best-fitting galaxy HOD for $\lcdm$ and the three DGP cases studied in this paper, as labelled. The left panels shows $N(M)$ (solid), $N_c(M)$ (dashed) and $N_s(M)$ (dotted) distributions, while the right panel shows these distributions multiplied by the halo mass function to show the relative contribution to the total galaxy number density from halos of a given mass. The panel inset on the right indicates where each best-fitting HOD model lies in $\left[\sigma_{{\rm log}_{10}M}, {\rm log}_{10}M_{\rm min}\right]$ space.}
\label{fig:hod}
\end{figure*}

\begin{table}
\caption{Summary of the cosmological models considered in this paper together with the respective true growth rate and best-fitting HOD effective linear galaxy bias at $z_{\rm CMASS} = 0.57$.}
\begin{tabular}{@{}lccccccccccc}
\hline\hline
\\
Model  & $r_cH_0$ & \ \ $f(z_{\rm CMASS})$ & $f\sigma_8(z_{\rm CMASS})$ & $b_g$ &\ \ 
\\
\hline
\\
$\lcdm$ &\ \  $\infty$ &  $0.77$ & $0.48$ & $1.95$ & \ \ 
\\
$\wdgp$ &\ \  $2.0$ &  $0.80$ & $0.51$ &  $1.88$ & \ \ 
\\
$\mdgp$ &\ \  $0.5$ &  $0.84$ & $0.56$ & $1.76$ & \ \ 
\\
$\sdgp$ &\ \  $0.1$ &  $0.89$ & $0.69$ & $1.51$ & \ \ 
\\
\hline
\hline
\end{tabular}
\label{table:models}
\end{table}

We construct our galaxy mock catalogues by populating the dark matter halos in the simulations within the halo occupation distribution (HOD) framework \cite{2004ApJ...609...35K}. In this framework, one parametrizes the mean number of galaxies that reside in halos of mass $M$, $N(M)$, and tunes the parameters such that the resulting galaxy distribution meets some desired observational constraints. We follow closely the HOD parametrization analysis of Refs.~\cite{2007ApJ...667..760Z, 2013MNRAS.428.1036M, 2016MNRAS.457.1577G} and split $N(M)$ into the sum of the number of central $N_c(M)$ and satellite $N_s(M)$ galaxies as 
\bq\label{eq:hodfunctions}
&&N(M) = N_c(M) + N_s(M), \\
&&N_c(M) = \nonumber \\
&&\frac{\Theta(M - M_{\rm res})}{2}\left[1 + {\rm erf}\left(\frac{{\rm log}_{10}M - {\rm log}_{10}M_{\rm min}}{\sigma_{{\rm log}_{10}M}}\right)\right], \\
&&N_s(M) = N_c(M)\left(\frac{M - M_0}{M_1'}\right)^\alpha.
\eq
The distribution of centrals is described by a smooth cut-off mass scale $M_{\rm min}$ and transition width $\sigma_{{\rm log}_{10}M}$ from the "no-central" to "one-central" regimes. {The factor $\Theta(M - M_{\rm res})$ is included to prevent centrals to be assigned to halos that fall below the mass resolution of our simulations $M_{\rm res} = 10^{12}\ M_{\odot}/h$ (this corresponds roughly to halos with 100 particles).} The satellite distribution is characterized by the cut-off scale $M_0$, normalization $M_1'$ and a power-law slope $\alpha$. We {choose} these HOD parameters to approximately match (i) the observed number density and (ii) the \emph{large-scale} amplitude of the power spectrum monopole of the CMASS LRG sample of the BOSS survey \cite{2014MNRAS.441...24A}. Before describing our fitting procedure below, we stress that the purpose of the HOD here is to have a physically reasonable N-body-based toy model of a real galaxy sample that resembles CMASS in terms of mean density and large-scale clustering amplitude.  We do not expect this HOD sample to match CMASS clustering on small scales $\lesssim 20 h^{-1}\,{\rm Mpc}$.  However, since we do not use those scales in the RSD analysis, any such mismatch is not relevant for the purposes of this paper (see discussion below).

The requirement (i) above dictates that
\bq\label{eq:const1}
n_g = \int{\rm d}M\frac{{\rm d}n}{{\rm d}M}N(M) = \bar{n}_{\rm CMASS} \approx 3.8 \times 10^{-4}\ h^3/{\rm Mpc}^3, \nonumber \\
\eq
where $n_g$ is the total galaxy number density, ${\rm d}n/{\rm d}M$ is the best-fitting ST mass function (cf.~Fig.~\ref{fig:cmf}) and $\bar{n}_{\rm CMASS}$ is the mean galaxy number density of the CMASS sample. To derive the expression for the requirement (ii) above, we relate the anisotropic galaxy power spectrum in redshift space $P_g^z(k, \mu)$ to the linear matter power spectrum in real space $P_{\rm lin}(k)$ as \cite{Kaiser:87}
\bq\label{eq:kaisermodel}
P_g^z(k, \mu) = b_g^2\left(1 + \beta\mu^2\right)^2 P_{\rm lin}(k),
\eq
where $\mu$ is the cosine of the angle between the line-of-sight and the wavevector $\vec{k}$, $k = |\vec{k}|$, $\beta = f/b_g$ and $b_g$ is the effective galaxy linear bias
\bq\label{eq:bg}
b_g = \frac{1}{n_g}\int{\rm d}M\frac{{\rm d}n}{{\rm d}M}N(M)b(M),
\eq
where $b(M)$ is the best-fitting ST linear halo bias shown in Fig.~\ref{fig:hb}. Equation (\ref{eq:kaisermodel}), which we assume is valid for both $\lcdm$ and DGP cosmologies, is expected to hold only on sufficiently large scales where nonlinearities in the matter distribution, nonlinear RSD and galaxy bias (as well as the scale dependence of bias) can be neglected. The galaxy power spectrum can be expanded in multipoles as
\bq
\label{eq:multi1}P_g^z(k, \mu) &=& \sum_\ell P_\ell^z(k)L_\ell(\mu), \\
\label{eq:multi2}P_\ell^z(k) &=& \frac{2\ell + 1}{2}\int_{-1}^1 P_g^z(k, \mu)L_\ell(\mu){\rm d}\mu,
\eq
where $L_{\ell}(\mu)$ are Legendre polynomials. Combining Eqs.~(\ref{eq:kaisermodel}) and (\ref{eq:multi2}), we have that the power spectrum monopole ($\ell = 0$; {i.e., angle average over the $\vec{k}$ direction}) is given by
\bq\label{eq:mono}
P_{\ell=0}^z(k) &=&  b_g^2\left[1 + \frac{2}{3}\beta + \frac{1}{5}\beta^2\right]P_{\rm lin}(k) \nonumber \\
&=& \mathcal{R} P_{\rm lin}^{\lcdm}(k),
\eq
where
\bq\label{eq:R}
\mathcal{R} = b_g^2\left[1 + \frac{2}{3}\beta + \frac{1}{5}\beta^2\right]\left(\frac{D}{D_{\lcdm}}\right)^2.
\eq
Here, $D$ denotes the linear growth factor normalized to $D(a) = a$ deep in the matter-dominated regime, at sufficiently early times so that the modified gravity effects are negligible.  Similarly, $D_{\lcdm}$ denotes the same for $\lcdm$.  
We have written Eq.~(\ref{eq:mono}) in terms of the linear matter power spectrum in the $\lcdm$ model to leave clear that in the DGP models one must take into account that the linear growth is also modified. For the HOD parameter values used in Ref.~\cite{2016MNRAS.457.1577G} we have that $\mathcal{R}= \bar{\mathcal{R}} \approx 4.97$ for $\lcdm$\footnote{Reference~\cite{2016MNRAS.457.1577G} finds $\bar{\mathcal{R}} \approx 5.15$.  The difference is due to a different fiducial cosmology.}. We take this $\bar{\mathcal{R}}$ value as the "target" to derive the observational requirement (ii) mentioned above.

In practice, we define the following "pseudo-$\chi^2$" quantities
\bq\label{eq:chi2s}
\chi^2_{n_g} &=& \left(n_g - \bar{n}_{\rm CMASS}\right)^2, \\
\chi^2_{\mathcal{R}} &=& \left(\mathcal{R} - \bar{\mathcal{R}}\right)^2,
\eq
and find the HOD parameters that minimize $\mathcal{P} \propto {\rm exp}\left[-\chi^2/2\right]$, with $\chi^2 = \chi^2_{n_g} + \chi^2_{\mathcal{R}}$.  We skip modelling the errors on the values of $\bar{n}_{\rm CMASS}$ and $\bar{\mathcal{R}}$ as we are mostly interested in the best-fitting HOD parameters, and less so in their uncertainties (hence the label "pseudo-$\chi^2$"). We fix the satellite HOD parameters to $M_0 = 10^{13.1}$, $M_1' = 10^{14.2}$ and $\alpha = 0.8$, as in Ref.~\cite{2016MNRAS.457.1577G}, {and vary only the central parameters $M_{\rm min}$ and $\sigma_{{\rm log}_{10}M}$. Note that since we have two constraints in Eqs.~(\ref{eq:chi2s}) for two free parameters, there is a unique solution with $\sum \chi^2 = 0$ that is independent of the normalization of the $\chi^2$ values.
} The satellite galaxy fraction is subdominant because they tend to live only in very high-mass halos, whose abundance is exponentially suppressed (cf.~Fig.~\ref{fig:cmf}). We checked explicitly that setting $N_s(M) = 0$ does not change the resulting central parameters appreciably.

The best-fitting HOD distributions are shown in the left panel Fig.~\ref{fig:hod}. The right panel shows the contribution of halos of a given mass to the galaxy number density. The values of the resulting effective galaxy bias are quoted in Table \ref{table:models}. Figure \ref{fig:hod} shows that the $N(M)$ distributions extend to lower halo mass values and become wider with increasing fifth force strength. This follows from a combination of effects triggered by the differences in halo abundances (cf.~Fig.~\ref{fig:cmf}), bias (cf.~Fig.~\ref{fig:hb}) and linear growth. To give an example, if galaxies populate lower mass halos, then this effectively reduces $b_g$ (cf.~Table \ref{table:models}) because lower mass halos are less biased, which helps to compensate the boosted amplitude of the linear matter power spectrum (note the degeneracy between $b_g$ and $D/D_{\lcdm}$ in Eq.~(\ref{eq:R})). Note also that for fixed halo mass, halos themselves are less biased in the DGP cosmologies than in $\lcdm$.  Still, at fixed mass the halo power spectrum is larger in DGP than in $\lcdm$, so that galaxies need to populate lower mass halos in the former model.  

The construction of the actual galaxy catalogues that go into the analysis described in the next section is as follows. A halo can either host one central or none and this is determined with probability $N_c(M)$. The position and peculiar velocity of the centrals is assigned to be that of its host halo. If a halo contains a central, then its number of satellites is drawn from a Poisson distribution with mean $N_s(M)$. The position and velocity of the satellites are assigned to be those of randomly chosen halo particles. We adopt the plane parallel approximation to "move" the galaxies from real to redshift space. 

One other possible way to assign galaxies to dark matter halos is via subhalo abundance matching (SHAM) methods \cite{2004MNRAS.353..189V, 2015arXiv150900482S}. In these, instead of parametrizing  the halo occupation number, one populates the halos by assuming that some set of observed galaxy properties (luminosity, stellar mass, etc.) are monotonically related to some halo property (mass, circular velocity, etc.). In practice, however, SHAM requires higher N-body resolution than that used here in order to properly resolve subhalos, which is why we shall limit our analysis to HOD-based catalogues (see e.g.~Ref.~\cite{2016MNRAS.457.1076J} for an assessment of the impact of using different methods to populate halos with galaxies on the clustering in redshift space). Furthermore, although we could have explored ways to determine the HOD parameters using clustering information on smaller scales (e.g., by using the projected galaxy correlation function \cite{2007ApJ...667..760Z, 2014MNRAS.444..476R}), we note that this is not crucial for our validation analysis of the clustering wedges model used in Refs.~\cite{2013MNRAS.433.1202S, 2014MNRAS.440.2692S} (see next section). The latter uses only clustering information on scales {$>20\ {\rm Mpc}/h$} (cf.~Fig.~\ref{fig:wedges} below), and as a result, it is sufficient to ensure that our $\lcdm$ and DGP mocks have similar large scale clustering properties (large scale monopole power spectrum in our case, cf.~Eqs.~(\ref{eq:mono}) and (\ref{eq:R})). Our mocks are therefore similar to those constructed in Ref.~\cite{2013MNRAS.428.1036M} (which have been used in covariance matrix estimations and other validation studies \cite{2013MNRAS.433.1202S, 2014MNRAS.440.2692S, 2016arXiv160703147S}), in that the HOD parameters are tuned to match some large (not small) scale clustering properties. We note also that a more elaborate fitting strategy based on smaller scale clustering would further require significantly higher resolution simulations and a more careful assignment of galaxy velocities \cite{2014MNRAS.444..476R}.

For simplicity, we also skip modelling the angular and redshift selection function of the CMASS sample (see e.g.~Ref.~\cite{2013MNRAS.428.1036M} for the steps towards that). This is not critical since our focus is on the impact of the different gravitational physics in the DGP and $\lcdm$ cosmologies, which is independent of the survey geometry and exact galaxy redshift distribution. 

\section{Growth rate estimates in modified gravity}\label{sec:main}

\subsection{Modelling RSD in the galaxy clustering statistics}\label{sec:method}

\begin{figure*}
	\centering
	\includegraphics[scale=0.41]{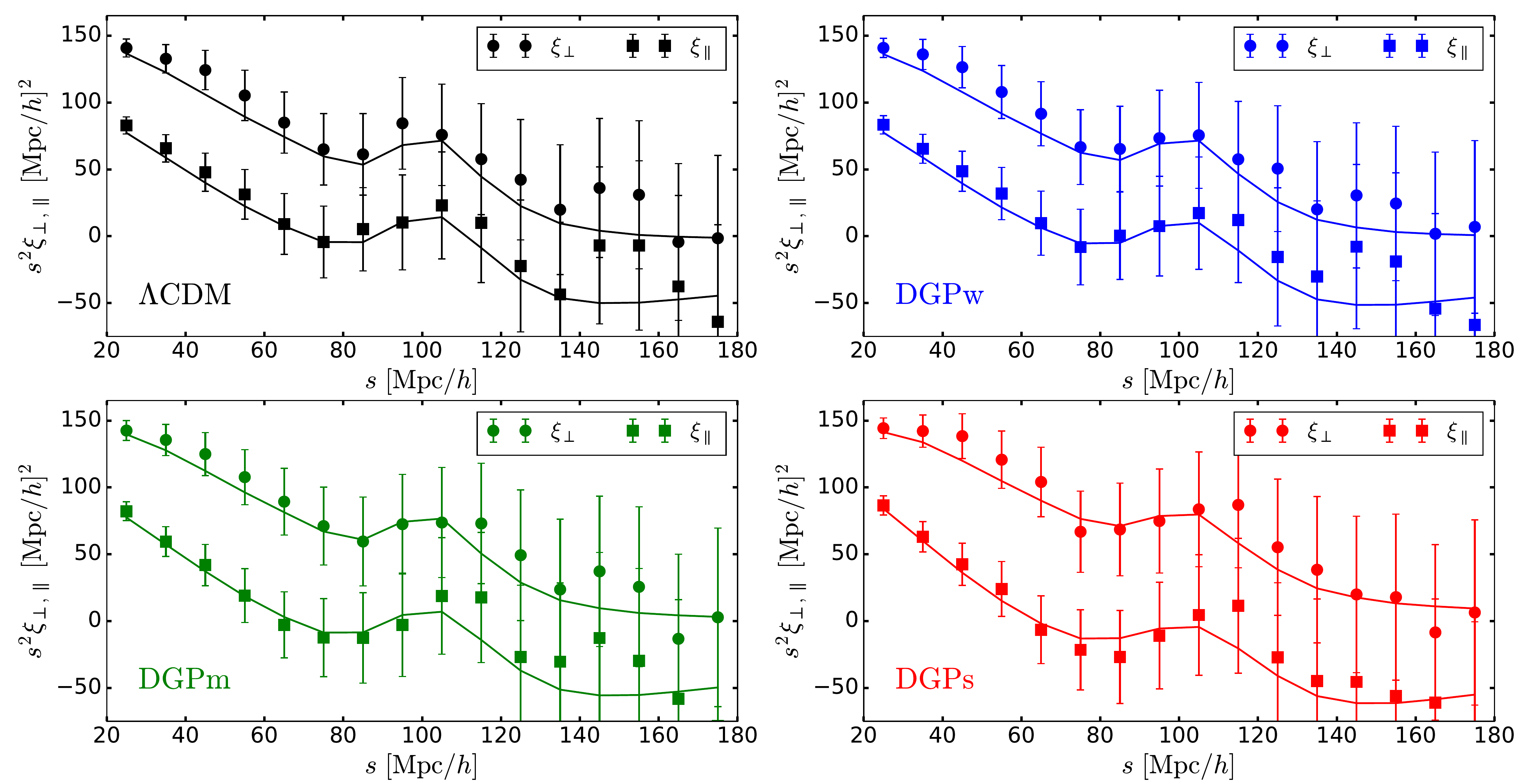}
	\caption{Clustering wedges $\xi_\perp$ (dots) and $\xi_\parallel$ (squares) measured from the mocks of $\lcdm$ and the three DGP cases, as labelled. The errorbars show the diagonal entries of the covariance matrix model of Ref.~\cite{2016MNRAS.457.1577G}. The solid lines show the best-fitting model outlined in Sec.~\ref{sec:method} {when the AP and $f\sigma_8$ parameters are fixed to their true values.}}
\label{fig:wedges}
\end{figure*}

We follow the recipe of Ref.~\cite{2016arXiv160703147S} to estimate the growth rate from our mocks by modelling the {\it clustering wedges} of the redshift-space galaxy two-point correlation function (2PCF), $\xi(s)$, where $s$ is the pair separation in redshift space.

\subsubsection{Clustering wedges}\label{sec:cwedges}

 The clustering wedges are defined as a projection of the 2PCF into bins of $\mu$
\bq\label{eq:wedges}
\xi_{\mu_1}^{\mu_2}(s) = \frac{1}{\mu_2 - \mu_1} \int_{\mu_1}^{\mu_2} \xi(s, \mu)\ {\rm d}\mu.
\eq
We consider two clustering wedges: $\xi_{\perp}$ for $\mu_1=0;\mu_2=0.5$, and $\xi_{\parallel}$ for $\mu_1=0.5;\mu_2=1$. Similarly to Eqs.~(\ref{eq:multi1}) and (\ref{eq:multi2}), one can expand $\xi(s, \mu)$ as
\bq\label{eq:2pcfexp}
\xi(s, \mu) = \sum_\ell \mathcal{L}_\ell(\mu)\xi_\ell(s),
\eq
with the multipole moments of the 2PCF $\xi_\ell(s)$ being related to those of the power spectrum by
\bq\label{eq:2pcfmultipk}
\xi_\ell(s) = \frac{i^\ell}{(2\pi)^3} \int_0^{\infty} P_\ell(k) j_\ell(ks){\rm d}^3k,
\eq
where $j_\ell(x)$ is the spherical Bessel function of order $\ell$. In Eq.~(\ref{eq:2pcfexp}), it suffices to consider only the monopole, quadropole and hexadecapole ($l = 0, 2, 4$, respectively), as higher multipoles give negligible contributions \cite{2013MNRAS.433.1202S}. By establishing a model for the galaxy power spectrum in redshift space, $P_g^z(k, \mu)$, one can get $P_\ell(k)$ using Eq.~(\ref{eq:multi2}) to then use Eqs.~(\ref{eq:wedges}), (\ref{eq:2pcfexp}) and (\ref{eq:2pcfmultipk}) to compute the clustering wedges and compare to those measured from the mocks.  

\subsubsection{Redshift galaxy power spectrum}\label{sec:galpk}

{We adopt the same modeling of $P_g^z(k, \mu)$ as in Ref.~\cite{2016arXiv160703147S}, which we describe very schematically below (we refer the interested reader to Ref.~\cite{2016arXiv160703147S} for the details).  We adopt the following parametrization
\bq\label{eq:Pkparam}
P_g^z(k, \mu) = W_{\infty}(i f k \mu)\sum_{i = 1}^3 P_{\rm novir}^{(i)}(k, \mu),
\eq
with
\bq\label{eq:genfunction}
W_{\infty}(\lambda) = \frac{1}{\sqrt{1 - \lambda^2a_{\rm vir}^2}} {\rm \exp}\left[\frac{\lambda^2\sigma_v^2}{1 - \lambda^2 a_{\rm vir}^2}\right],
\eq
where $a_{\rm vir}$ is a free parameter and $\sigma_v = \left[I_0(0) + I_2(0)\right]/3$, with $I_l(r) = \int {\rm d}^3kj_{l}(kr)P_{\rm lin}(k)/k^2$. The function $W_{\infty}$ models the suppression of the clustering power on small scales, also commonly referred to as the "fingers-of-god" effect. The remaining three $P^{(i)}$ terms are given in terms of the real-space galaxy density or galaxy velocity power spectra and bispectra. For instance, 
\bq\label{eq:Pi1}
P^{(1)}_{\rm novir}(k, \mu) = P_{g}(k) + 2f\mu^2P_{g\theta}(k) + f^2\mu^4P_{\theta}(k),
\eq
where $P_g$ and $P_{\theta}$ are, respectively, the power spectrum of the galaxy density contrast $\delta_g$ and galaxy velocity divergence $\theta$. Their cross spectrum is $P_{g\theta}$. We refer the reader to Ref.~\cite{2016arXiv160703147S} for the expressions of $P^{(2)}$ and $P^{(3)}$ and for details about how they are evaluated.}

{The calculation of the above $P_{\rm novir}^{(i)}$ terms requires a model for the matter and velocity power 
spectra on mildly nonlinear scales. In the model we test in this paper, these are given by 
an extension of the idea behind renormalized perturbation theory (RPT, \cite{2006PhRvD..73f3519C}) dubbed 
gRPT \cite{inprep1}.
According to RPT, $P_{\rm NL}$ can be written as (see also Refs.~\cite{2008PhRvD..78j3521B, 2012arXiv1211.1571B, 2012PhRvD..85l3519B, 2012PhRvD..86j3528T, 2013PhRvD..87h3509T})
\bq
P_{\rm NL}(k) = P_{\rm lin}(k)G(k)^2 + P_{\rm MC}(k),
\eq
where $G(k)$ corresponds to a resummation of all the terms in the perturbative expansion that are proportional to 
the linear matter power spectrum and $P_{\rm MC}(k)$ includes mode-coupling terms (convolutions of linear spectra). 
In gRPT, Galilean invariance is used to find a resummation of the mode-coupling power consistent with the 
resummation of the propagator, providing an improved description of $P_{\rm NL}$ down to smaller scales.}

{The last ingredient that goes into the calculation of $P_g^z(k, \mu)$ is a galaxy bias model to relate $\delta_g$ to $\delta$. The model we use sets this relation as \cite{2012PhRvD..85h3509C, 1996ApJ...461L..65F, 1998MNRAS.297..692C}
\bq\label{eq:biasmodel}
\delta_g = &&b_1\delta + \frac{b_2}{2}\delta^2 + \gamma_2\mathcal{G}_2(\bar{\Phi}_v) + \gamma_3^-\left[\mathcal{G}_2(\bar{\Phi}) - \mathcal{G}_2(\bar{\Phi}_v)\right] \nonumber \\
\eq
where $\mathcal{G}_2(x) = \left[\left(\partial_i \partial_j x\right)^2 - \left(\nabla^2x\right)^2\right]$, with $\bar{\Phi}$ and $\bar{\Phi}_v$ defined as $\nabla^2\bar{\Phi} = \delta$ and $\nabla^2\bar{\Phi}_v = \theta_m$ (with ${\theta_m}$ being the velocity divergence of the matter field). As in Ref.~\cite{2016arXiv160703147S}, we use the relation $\gamma_2 = -2(b_1 - 1)/7$, which results in a galaxy bias model with three free parameters.}

{We have also tested the performance of the simpler $P_g^z(k, \mu)$ modelling used in the previous DR10 and DR11 BOSS clustering wedges analysis \cite{2013MNRAS.433.1202S, 2014MNRAS.440.2692S}. We have found that our conclusions for DR12 hold also in the DR11 and DR10 cases. In this paper, for brevity, we limit ourselves to showing only the results for the model outlined in this subsection since it is that which was used in the final data release from BOSS (the interested reader can find a validation analysis of the DR10/DR11 RSD model in version 1 and 2 of this manuscript at http://arxiv.org/abs/1605.03965)}.

\subsubsection{The Alcock-Paczynski effect}\label{sec:APeffect}

In the analysis of real galaxy surveys one has to assume a fiducial cosmology to convert the measured redshifts into distances. Additional anisotropies in the clustering pattern are then introduced if the fiducial cosmology is not the true one, which is known as the Alcock-Paczynski (AP) effect. In order to get unbiased constraints on $f$ from real galaxy samples, one must therefore include the AP effect into the RSD model. This can be done by analysing the data with some fiducial cosmology and introduce the AP parameters
\bq\label{eq:APparams}
\alpha_\perp = \frac{d_A(z)}{\tilde{d}_A(z)}\ \ ;\ \ \alpha_\parallel = \frac{\tilde{H}(z)}{H(z)},
\eq
where $d_A$ is the true angular diameter distance and a tilde denotes a quantity calculated in the fiducial cosmology. Distances and cosine angle scales in some cosmology $(s, \mu)$ are related to those in the fiducial one $(\tilde s, \tilde\mu)$ as
\bq\label{eq:smurelations}
s &=&  \tilde s\sqrt{\alpha_\parallel^2 \tilde\mu^2 + \alpha_\perp^2(1 - \tilde \mu^2)}, \\
\mu &=& \tilde\mu \frac{\alpha_\parallel}{\sqrt{\alpha_\parallel^2 \tilde \mu^2 + \alpha_\perp^2(1 - \tilde \mu^2)}}.
\eq
Deviations of $\alpha_\perp$ and $\alpha_\parallel$ from unity provide a measure for how much different the fiducial cosmology is from the true (unknown) one, and can be used to place constraints on the angular diameter distance and Hubble rate at the mean redshift of the galaxy samples analysed.

\subsubsection{Treatment of free parameters}\label{sec:nuisance}

{In linear theory, there is a well known degeneracy between $b_1$, $f$ and the amplitude of the linear matter power spectrum. The latter is normally described by the parameter $\sigma_8$, which is the {\it root mean squared} fluctuation of the linear density field on $8\ {\rm Mpc}/h$ scales. This degeneracy still holds to a significant degree on mildly nonlinear scales, which is why RSD constraints are typically phrased in terms of the parameter combination $f\sigma_8$, which absorbs the normalization of the linear power spectrum \cite{2009JCAP...10..004S}. As for the shape of $P_{\rm lin}$ (which enters as an ingredient of the RSD model), we take it to be that which corresponds to the input cosmological parameters of our simulations. Naturally, in real observational analysis these parameters are also fitted against the data. Here, however, in order to improve our constraints on the remaining parameters of the RSD model, we shall assume perfect knowledge of the shape of $P_{\rm lin}$, which is the same for $\lcdm$ and DGP gravity up to an overall normalization captured by $\sigma_8$.}

{In total, the RSD model described above contains seven free parameters, $f\sigma_8$, $b_1$, $b_2$, $\gamma_3^-$, $a_{\rm vir}$, $\alpha_\perp$ and $\alpha_\parallel$, which we can constrain by fitting them against the clustering wedges $\xi_\perp$ and $\xi_\parallel$ measured from the mocks. In particular, in this paper we are interested in determining if the constraints on $f\sigma_8$ are consistent with its expected values. In order to improve the statistical uncertainty on the estimated $f\sigma_8$ from the mocks, we shall fix the AP parameters to unity and take the background cosmology of the mocks as the fiducial one. Recall that with the same purpose we have also assumed perfect knowledge of the shape of the linear matter power spectrum. We stress that relaxing these assumptions would only deteriorate the statistical significance of our constraints, without affecting them in a systematic way. In our fitting analysis, we use the Gaussian covariance matrix model presented in Ref.~\cite{2016MNRAS.457.1577G} to account for statistical errors in our mock measurements. Before proceeding, note that the RSD model does not have any ingredient that aims to capture the impact of any specific modified gravity feature (we shall return to the discussion of this point below).}

\subsection{Growth rate estimates from the mocks}\label{sec:results}

\begin{figure*}
	\centering
	\includegraphics[scale=0.42]{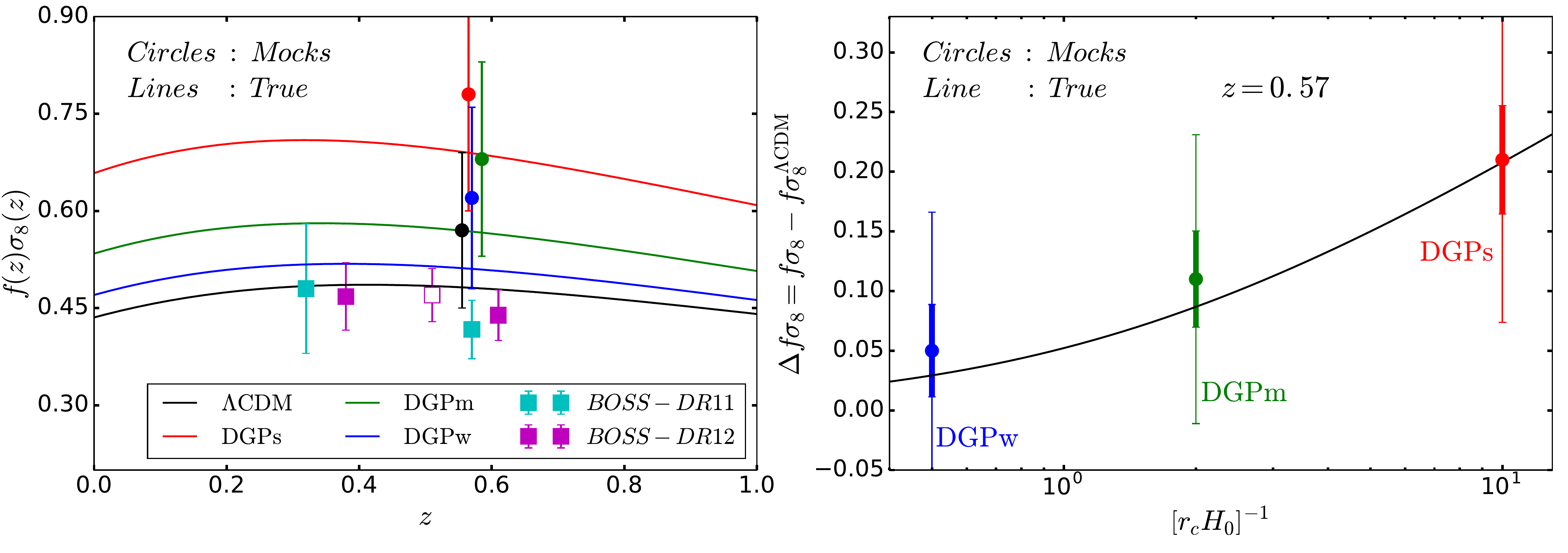}
	\caption{{Growth rate estimates from the analysis of the galaxy mock catalogues of $\lcdm$ and the three DGP cases, as labelled. The left panel shows the expected time evolution of $f\sigma_8$ (lines) together with the corresponding estimates obtained from the mocks (dots with errorbars; all points correspond to $z=0.57$, but have been slightly shifted for better visualization). The purple squares correspond to the estimates obtained in Ref.~\cite{2016arXiv160703147S} from a similar analysis applied to the DR12 galaxy sample from BOSS in three redshift bins, $z = 0.38, 0.51, 0.61$ (the open symbol used for the mid-redshift point is used to remind that this point is very covariant with the other two). The cyan squares show the estimates obtained in Ref.~\cite{2014MNRAS.440.2692S} using a different RSD clustering wedges model applied to the LOWZ ($z = 0.32$) and CMASS ($z = 0.57$) BOSS galaxy samples from DR11. The right panel shows the expected difference to $\lcdm$ in $f\sigma_8$ as a function of $\left[r_cH_0\right]^{-1}$ for $z = 0.57$ (line), together with the difference measured from the mocks. The two errorbar sizes for each DGP model correspond to two different values of $T_{\rm CV}$ in Eq.~(\ref{eq:errorscale}).}}
\label{fig:fs8fit}
\end{figure*}

The clustering wedges $\xi_{\perp}$ and $\xi_{\parallel}$ measured from our HOD mocks are shown in Fig.~\ref{fig:wedges} (symbols with errorbars). In each panel, the result corresponds to the average clustering wedges over three catalogues, each obtained by employing the plane parallel approximation using one of the three cartesian axes of the simulation box. {The solid lines show the resulting best-fitting clustering wedges model obtained from constraints in which the AP and $f\sigma_8$ parameters are fixed to their true values.} The figure shows that the RSD model provides a good description of the clustering in our $\lcdm$ and DGP gravity mocks. The left panel of Fig.~\ref{fig:fs8fit} shows the constraints on $f\sigma_8$ obtained from the mocks (dots with errorbars). {We see that, although the bounds on the $f\sigma_8$ constraints are compatible with the corresponding true values at $z = 0.57$, the mean of the constraints is markedly overpredicting them.} The errorbars are also rather large in comparison to other analyses of CMASS-like mocks, which is mostly due to the smaller size of our simulation boxes (see e.g.~Ref.~\cite{2016MNRAS.457.1577G}, in which $L_{\rm box} = 1500 {\rm Mpc}/h$). 

{The absolute value of $f\sigma_8$ is not the best quantity to compare the theoretical expectation with because we only have one realization of the initial conditions, and hence, the comparison is heavily affected by cosmic variance. To overcome this, we focus instead on the difference in $f\sigma_8$ between DGP and $\lcdm$, $\Delta f\sigma_8 = f\sigma_8^{\rm DGP} - f\sigma_8^{\lcdm}$. Given that this RSD model has been shown to retrieve unbiased constraints of $f\sigma_8$ in $\lcdm$ mocks when many realizations are available \cite{2016arXiv160703147S}, if $\Delta f\sigma_8$ for a single realization is unbiased, then we can expect unbiased constraints on $f\sigma_8$ in DGP gravity as well, if more realizations of the simulations are made available too. Another advantage of focusing on $\Delta f\sigma_8$ is that its errorbar can be scaled down by appropriately taking into account the fact that our simulations evolved from the same initial conditions (cf.~Appendix \ref{sec:cv}).} 

The expected result of $\Delta f\sigma_8$ at $z = 0.57$ is shown in the right panel of Fig.~\ref{fig:fs8fit} as a function of $\left[r_cH_0\right]^{-1}$, together with the values estimated from the mocks. The errorbars on $\Delta f\sigma_8$ are given by
\bq\label{eq:errorscale}
E\left[\Delta f\sigma_8\right] = T_{\rm CV} \sqrt{E\left[f\sigma_8^{\rm DGP}\right]^2 + E\left[f\sigma_8^{\lcdm}\right]^2},
\eq
where $E\left[a\right]$ denotes the error on quantity $a$ and $T_{\rm CV}$ is a reduction factor that arises because we are comparing simulation results that evolved from the same initial conditions. The two errorbars shown for each DGP model correspond to a conservative and our best estimate of the reduction factor, { $T_{\rm CV} = 0.63$ and $T_{\rm CV} = 0.21$}, respectively. These factors are derived in Appendix \ref{sec:cv}. We note also that since our clustering wedges results correspond to an average over the result obtained by taking each of the three simulation axes as the line of sight direction, our {\it effective} volume is actually larger than $V=600^3 {\rm Mpc^3}/h^3$. Naturally, these different clustering measurements are not independent, but one expects nevertheless the errorbars to be scaled down further due to this. We opted to remain conservative and not include an estimate for this effect.

From Fig.~\ref{fig:fs8fit} we therefore conclude that, {\it within the statistical precision attained by our simulations, we find no evidence that the clustering wedges model yields biased estimates of the growth rate in DGP cosmologies.} This is the main result of this paper, which holds even for cases where $f \sigma_8$ is enhanced by more than $40\%$, relative to $\lcdm$.

It is interesting to link our results with those of Ref.~\cite{2014PhRvD..89d3509T}. There, the authors find that RSD models can yield biased constraints when applied to $f(R)$ gravity cosmologies, unless the former are appropriately generalized to include modified gravity effects. On the other hand, the RSD model we use contains no explicit ingredient for modified gravity, and yet we find that the same model is as successful in DGP cosmologies as it is in $\lcdm$. There are a few differences in our analysis and that of Ref.~\cite{2014PhRvD..89d3509T} that make a direct comparison of the results difficult, namely (i) the RSD models are not the same; and (ii) we apply ours on galaxy HOD catalogues, whereas in  Ref.~\cite{2014PhRvD..89d3509T} the authors apply theirs on simulation snapshots of the whole mass distribution.  Note that the redshift-space distortion of the matter density field is much stronger than that of the halo and mock galaxy fields.  Another important difference in these two works lies in the phenomenology of the two theories of gravity tested. In case of $f(R)$, the growth of structure is scale dependent even on linear scales, and the modified gravity effects disappear on large scales;  the latter fact significantly increases the importance of small scales for the constraints on $f(R)$.  Both of these do not hold in DGP gravity. As a result, it may be reasonable to assume that RSD models that work successfully in $\lcdm$ are less biased when applied to modified gravity cosmologies if the growth of structure on large scales remains scale independent.  One might therefore expect that our conclusions extend beyond DGP gravity and should hold in any theory with scale-independent linear growth. Along similar lines, one might reasonably expect that our conclusions also apply to other methods (for example power spectrum multipoles) to extract $f\sigma_8$ from large-scale clustering data \cite{2012MNRAS.423.3430B, 2012MNRAS.425..405B, 2013A&A...557A..54D, 2014MNRAS.439.2515O, 2014MNRAS.439.3504S}. A more detailed investigation of these considerations is beyond the scope of this paper. 

\subsection{Growth rate constraints on DGP gravity}\label{sec:constraints}

\begin{figure*}
	\centering
	\includegraphics[scale=0.41]{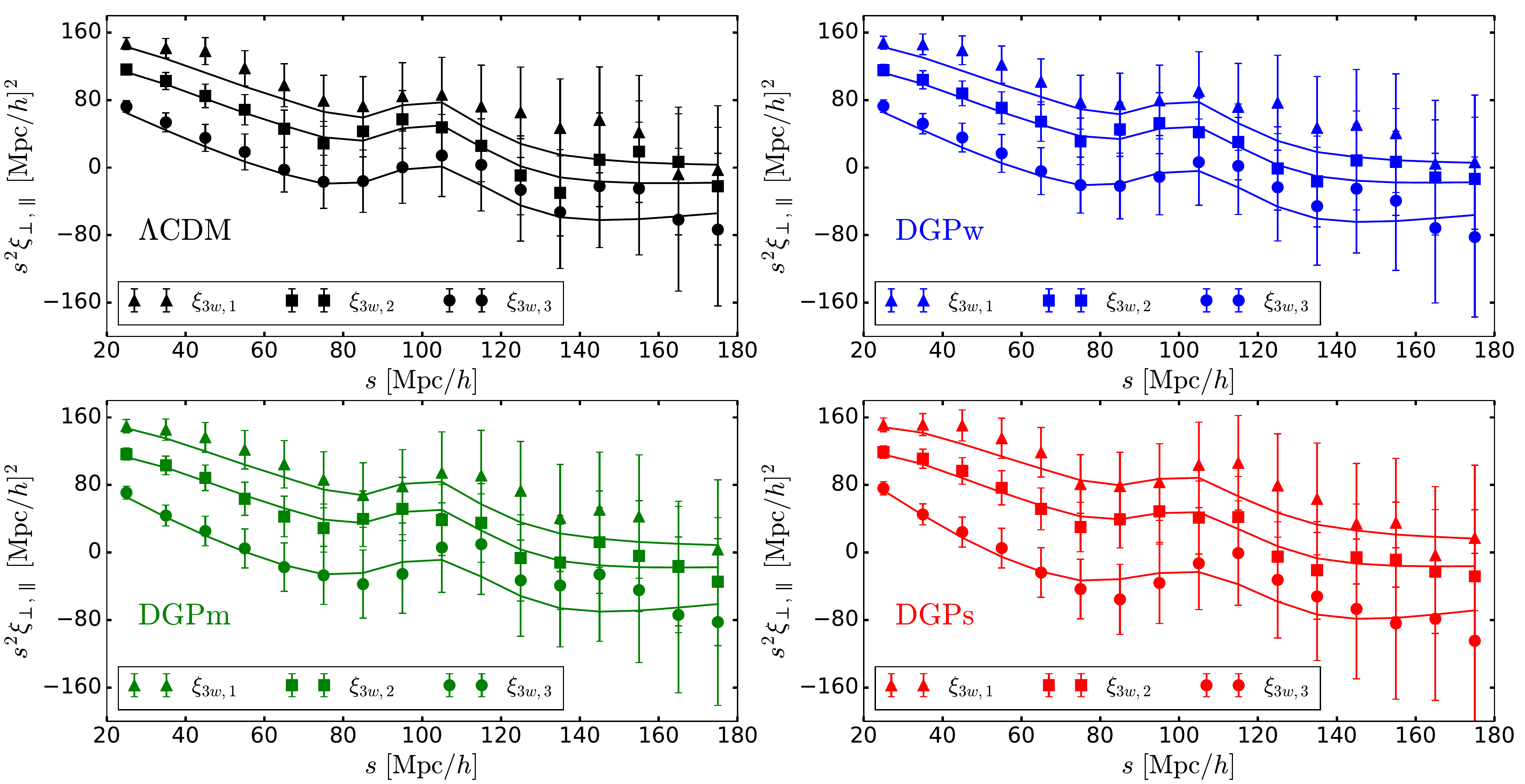}
	\caption{{Clustering wedges $\xi_{3w, i}$ $i=1,2,3$ measured from the mocks of $\lcdm$ and the three DGP cases, as labelled. The solid lines show the prediction of the best-fitting model outlined in Sec.~\ref{sec:method} constrained using the two wedges $\xi_\perp$ and $\xi_{\parallel}$.}}
\label{fig:3wedges}
\end{figure*}

\begin{figure}
	\centering
	\includegraphics[scale=0.42]{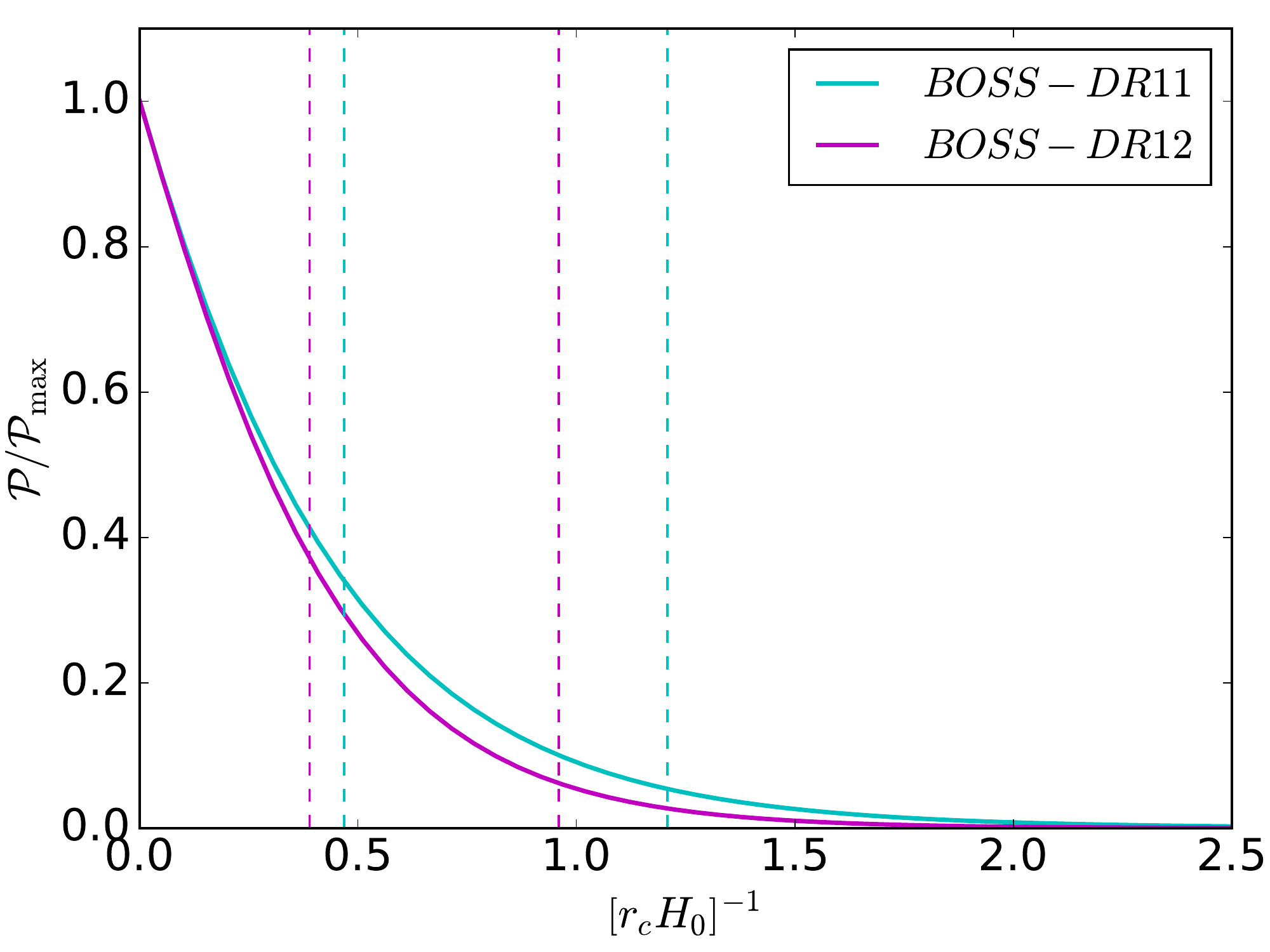}
	\caption{{Observational constraints on the DGP parameter $r_cH_0$ using the $f\sigma_8$ values obtained from the analysis of the BOSS DR12 (purple squares in Fig.~\ref{fig:fs8fit}) and DR11 (cyan squares in Fig.~\ref{fig:fs8fit}) data, as labelled. The results assume the priors $\Omega_{m0} = 0.315 \pm 0.013$ ($1\sigma$) and $\sigma_8^{\Lambda{\rm CDM}}(z=0) = 0.829 \pm 0.014$ ($1\sigma$), from Planck \cite{2015arXiv150201589P}. The lines shows the resulting probability distributions after marginalizing over $\Omega_{m0}$ and $\sigma_8^{\lcdm}$. The dotted vertical lines indicate the $68\%$ and $95\%$ percentiles of the distributions.}} 
\label{fig:P}
\end{figure}

{The successful application of the clustering wedges model to our mocks justifies using the observational estimates of $f\sigma_8$ obtained in the real data analysis of Ref.~\cite{2016arXiv160703147S} with the same RSD model to constrain DGP cosmologies. Before proceeding though, there is a difference between the analysis displayed in this paper and that of Ref.~\cite{2016arXiv160703147S} that is worth mentioning. Here, we have constrained $f\sigma_8$ from the mocks using two clustering wedges $\xi_{\perp}, \xi_{\parallel}$, whereas the observational analysis of Ref.~\cite{2016arXiv160703147S} makes use of three wedges, $\xi_{3w, i}$, characterized by $\left(i-1\right)/3 < \mu < i/3$, with $i=1,2,3$. One could therefore wonder whether the unbiased performance of the model holds also in the three wedge case. 
{The measurements of three wedges are more sensitive to the hexadecapole of the two-dimensional
correlation function, $\xi_4(s)$. Given the small volume of our simulations, $\xi_4(s)$ is dominated by noise, 
which could affect our constraints. For this reason we do not attempt to constrain $f\sigma_8$ using the three 
wedges measured from our mocks directly.
Instead, we check }the consistency between the measured three wedges with the corresponding prediction of the 
best-fitting model to the 
two wedges. This comparison is shown in Fig.~\ref{fig:3wedges}. The figure shows that the best-fitting model constrained with two wedges provides a reasonable description of the amplitude and shape of the three wedges as well (even if it was not fitted to them).  {While we cannot obtain a precise goodness-of-fit value due to the lack of a robust covariance for our small-volume mocks, conservative estimates yield a reduced $\chi^2$ of the fits in Fig.~\ref{fig:3wedges} that is less than 1.5.}  
This leads us to conclude that the performance of the RSD model is not dependent on the use of two or three clustering wedges. Furthermore, although we have only explicitly demonstrated the validity of the clustering wedges model at $z = 0.57$, we expect our conclusions to hold for other redshift values as well.}

{To constrain the cross-over scale $r_c$, we use the following DR12 $f\sigma_8$ estimates (purple squares in Fig.~\ref{fig:fs8fit}): $f\sigma_8(z = 0.38) = 0.468 \pm 0.052$, $f\sigma_8(z = 0.51) = 0.470 \pm 0.041$ and $f\sigma_8(z = 0.61) = 0.439 \pm 0.039$ (cf. Table 4 of Ref.~\cite{2016arXiv160703147S}). {The DR12 constraint on $f\sigma_8(z=0.51)$ is strongly correlated with those at lower and higher redshifts and does not lead to a significant improvement in the constraining power of these measurements.  In the results presented below, we use only the information from the estimates at $z = 0.38$ and $z = 0.61$, which we treat as independent.} For comparison, we consider also the previous DR11 LOWZ and CMASS sample results of Ref.~\cite{2014MNRAS.440.2692S}, which finds $f\sigma_8^{\rm LOWZ}(z = 0.3) = 0.48 \pm 0.1$ and $f\sigma_8^{\rm CMASS}(z = 0.57) = 0.417 \pm 0.045$ (cyan squares in the left panel of Fig.~\ref{fig:fs8fit}).} We sample the following three-dimensional parameter space: $\left\{\left[r_cH_0\right]^{-1}, \Omega_{m0}, \sigma_8^{\lcdm}(z=0)\right\}$.  Here, $\sigma_8^{\lcdm}$ is essentially a proxy for the primordial amplitude $A_s$ of scalar perturbations.  For a given point in this parameter space, we evaluate $\sigma_8(z)$ in the DGP model from its corresponding present-day value in $\lcdm$ $\sigma_8(z=0)^{\lcdm}$ as
\bq\label{eq:s8dgp}
\sigma_8^{\rm DGP}(z) = \frac{D^{\rm DGP}(z)}{D^{\lcdm}(z=0)}   \sigma_8^{\lcdm}(z=0).
\eq
The values of $f$ are obtained by solving Eq.~(\ref{eq:linearD}) for every given value of $\left[r_cH_0\right]^{-1}$ and $\Omega_{m0}$.

We consider also Gaussian priors with $\Omega_{m0} = 0.315 \pm 0.013$ ($1\sigma$) and $\sigma_8^{\Lambda{\rm CDM}} = 0.829 \pm 0.014$ ($1\sigma$), which correspond to Planck constraints assuming flat $\lcdm$ \cite{2015arXiv150201589P}. One may wonder whether the Planck bounds on $\Omega_{m0}$ obtained assuming $\lcdm$ models are the same if one had assumed a DGP cosmology instead. To address this point, we recall that these two models are indistinguishable at the time of recombination and have the same expansion history at all times. Hence, they predict the same high-$\ell$ CMB temperature spectra $\ell \gtrsim 50$. The only difference between the two models is the late-time evolution of the density fluctuations, {which can lead to different ISW and CMB lensing signals.} The ISW effect, however, is only important on the largest angular scales of the CMB spectra, on which cosmic variance does not permit great constraining power. {In contrast, the CMB lensing potential power spectrum data has more constraining power and one could wonder whether the bounds on $\Omega_m$ could display some degeneracies with $r_cH_0$. A full CMB constrain analysis is beyond the scope of the present paper. For simplicitly then, we opt to straightforwardly use the Planck bounds obtained assuming $\lcdm$ in Ref.~\cite{2015arXiv150201589P}, but keeping in mind that the resulting bounds on $r_cH_0$ depend on the assumed priors of $\Omega_m$. These considerations can be extended also to our assumption of a vanishing curvature energy density, $\Omega_K = 0$.}

{The marginalized constraints on $\left[r_cH_0\right]^{-1}$ using the DR11 and DR12 data points are shown in Fig.~\ref{fig:P} by the solid lines, as labelled. The vertical dashed lines indicate the $1\sigma$ and $2\sigma$ bounds. The figure illustrates that the constraints obtained using the DR12 data yield slightly tighter constraints on $\left[r_cH_0\right]^{-1}$. In particular, the DR12 points constrain $\left[r_cH_0\right]^{-1} < 0.97\ (2\sigma)$, whereas for DR11 we have $\left[r_cH_0\right]^{-1} < 1.2\ (2\sigma)$. For both DR11 and DR12, our $\sdgp$ ($\left[r_cH_0\right]^{-1} = 10$) and $\mdgp$ ($\left[r_cH_0\right]^{-1} = 2$) cases are in fact in severe tension with the data. Only the $\wdgp$ case ($\left[r_cH_0\right]^{-1} = 0.5$) lies in a region where the likelihood is sizeable. Note that $\lcdm$ corresponds to $\left[r_cH_0\right]^{-1} = 0$.}

Our constraint result can be compared to that obtained in Ref.~\cite{2013MNRAS.436...89R} by using measurements of the monopole and quadrupole of the 2PCF of the LRG sample from the DR7 of SDSS II. There, the authors find that $r_c > 340\ {\rm Mpc}$ ($2\sigma$), which translates into\footnote{We assumed $H_0 = 67.31\ {\rm km/s/Mpc}$ and have appropriately added the speed of light $c$ in this conversion. Explicitly, one has $\left[r_cH_0/c\right]^{-1} < 13.1$ ($2\sigma$), but recall we quote results with units where $c=1$. Alternatively, our constraint can also be quoted as $r_c > 2500\ {\rm Mpc}/h$ or $r_c > 3700\ {\rm Mpc}$.} $\left[r_cH_0\right]^{-1} < 13.1$ ($2\sigma$). The constraints from the DR11 and DR12 $f\sigma_8$ estimates presented here lead to more than an order of magnitude improvement. References \cite{2009PhRvD..80f3536L, 2010PhRvD..82d4032W, 2014JCAP...02..048X} have also placed observational constraints on the normal branch of the DGP model, but their results cannot be directly compared to ours because of different assumptions about the background evolution (in these references, the dark energy term in Eq.~(\ref{eq:dgpH}) is taken to evolve as the cosmological constant, $\Omega_{\rm de} = \Omega_{\Lambda}$).

\section{Summary and Conclusions}\label{sec:conc}

We have investigated the performance of RSD models of the clustering wedges of the galaxy two-point correlation function in recovering the true growth rate of structure $f\sigma_8$ in modified gravity cosmologies. Models of RSD are in general thoroughly validated against cosmological simulations with GR, but hardly in cases with alternative theories of gravity. As a result, it is imperative to perform validation tests in modified gravity scenarios in order to determine whether or not the estimated values of $f\sigma_8$ are biased, and therefore, whether they can be used directly to constrain modified gravity theories.

To do so, we have run cosmological simulations of $\lcdm$ and the normal branch of DGP gravity with $\lcdm$ background, which we have used to construct CMASS-like galaxy mock catalogues using a HOD framework (cf.~Sec.~\ref{sec:hodmodel}). We have considered three parameter values for the cross-over scale parameter of DGP gravity: $r_cH_0 = 0.1$, $r_cH_0 = 0.5$ and $r_cH_0 = 2.0$. We have called these three cases $\sdgp$, $\mdgp$ and $\wdgp$, respectively. We tuned the HOD model parameters such that the resulting catalogues approximately match the galaxy number density and large scale amplitude of the power spectrum monopole of the CMASS sample from the BOSS survey. The three DGP cases and $\lcdm$ have different best-fitting halo occupation distributions (cf.~Fig.~\ref{fig:hod}), which is expected given that they also have different halo abundance, linear halo bias (cf.~Figs.~\ref{fig:cmf} and \ref{fig:hb}) and linear matter power spectrum amplitude. We estimate $f\sigma_8$ from the mocks using the clustering wedges RSD model {used in the BOSS DR12 analysis of Ref.~\cite{2016arXiv160703147S}, which has shown to return unbiased constraints when applied to a suite of $\lcdm$ mocks. Although not shown in this paper (see versions 1 or 2 of this manuscript at http://arxiv.org/abs/1605.03965), we have also tested the simpler RSD model used in the previous BOSS DR10 and DR11 analyses of Refs.~\cite{2013MNRAS.433.1202S, 2014MNRAS.440.2692S}, finding the same conclusions.}

Our main result from applying the RSD model to the mocks is that we find no evidence for a bias in the estimated value for $\Delta f\sigma_8$, defined as the difference in $f\sigma_8$ between DGP and $\lcdm$ (cf.~right panel of Fig.~\ref{fig:fs8fit}). The absolute value of $f\sigma_8$ estimated from the mocks is not the best quantity to compare the theoretical expectation with because we only have one realization of the initial conditions (and hence the result is prone to cosmic variance effects). The focus on $\Delta f\sigma_8$ also allows us to improve the statistics of our constraints by taking into account the fact that the simulations evolved from the same initial conditions (cf.~Appendix \ref{sec:cv}). The unbiased performance of the RSD model when applied to our DGP mocks indicates that it is safe to constrain this theory of gravity using the estimates of $f\sigma_8$ from real data obtained with the same RSD analysis pipeline. {We used the DR12 $f\sigma_8$ estimates of Ref.~\cite{2016arXiv160703147S} to constrain $\left[r_cH_0\right]^{-1} < 0.97$ ($2\sigma$, cf.~Fig.~\ref{fig:P}), after marginalizing over $\Omega_{m0}$ and $\sigma_8^{\lcdm}(z=0)$ with Planck priors (cf.~Sec.~\ref{sec:constraints}). Using the previous DR11 LOWZ and CMASS $f\sigma_8$ estimates yields a slightly looser bound, $\left[r_cH_0\right]^{-1} < 1.2$ ($2\sigma$). These constraints represent more than an order of magnitude improvement over previous constraints on the normal branch of DGP gravity with a $\lcdm$ background.  
In the context of so-called \emph{self-accelerating} models that lead to acceleration without the need for Dark Energy, one expects the natural value of the cross-over scale to be of order $r_c \sim H_0^{-1}$, as is the case in the original self-accelerating DGP model \cite{DeffayetEtal02}.  Thus, our constraints are cutting into the most interesting region of the parameter space.  Further, future constraints on $r_c$ that limit this scale to values much smaller than $H_0^{-1}$ can be seen as pushing such models into a fine-tuned region of parameter space.  Note however that the baseline model considered here is not self-accelerating in any case, and can only serve as a toy model for viable future self-accelerating models with Vainshtein-type screening.}

In Ref.~\cite{2014PhRvD..89d3509T}, with the aid of N-body simulations of $f(R)$ cosmologies, the authors found that if modified gravity effects are not explicitly included in RSD modelling, then this can lead to biased constraints. This is a result that is in apparent contrast with the fact that the RSD model used here (which does not include any modelling of modified gravity) returns unbiased constraints when applied to the DGP mocks. Even though there are some important differences between the RSD models used here and in Ref.~\cite{2014PhRvD..89d3509T}, we believe that the origin of the apparently distinct conclusions is mainly associated with differences in the phenomenology of the two theories of gravity tested. In particular, in DGP gravity, the linear growth of structure is scale-independent (as in GR, but with a time-dependent $G_{\rm eff}$), which is an effect that may be more easily absorbed by the nuisance parameters of a RSD model. On the other hand, the linear growth of structure is manifestly scale-dependent in $f(R)$, which may imply the need to generalize existing RSD models to account for the richer phenomenology in these gravity scenarios. Based on these considerations, we expect our results to be valid for other modified gravity models (not just DGP), so long as their linear growth is scale-independent and they employ a screening mechanism of the Vainshtein type. A definite answer however can only be reached by repeating the analysis presented in this paper for these other theories of gravity.

\begin{acknowledgments}

We thank Carlton Baugh, Martin Crocce, Wojciech Hellwing, Elise Jennings, Baojiu Li, Shun Saito, Roman Scoccimarro and Difu Shi for very useful comments and discussions.  AGS acknowledges support from the Trans-regional Collaborative Research Centre TR33 ``The Dark Universe'' of the German Research Foundation (DFG). FS acknowledges support from the Marie Curie Career Integration Grant  (FP7-PEOPLE-2013-CIG) ``FundPhysicsAndLSS.'' This work used the DiRAC Data Centric system at Durham University, operated by the Institute for Computational Cosmology on behalf of the STFC DiRAC HPC Facility (www.dirac.ac.uk). This equipment was funded by BIS National E-infrastructure capital grant ST/K00042X/1, STFC capital grant ST/H008519/1, and STFC DiRAC Operations grant ST/K003267/1 and Durham University. DiRAC is part of the National E-Infrastructure.

\end{acknowledgments}

\appendix

\section{Best-fitting Sheth-Tormen mass function}\label{sec:stfit}

\begin{figure}
	\centering
	\includegraphics[scale=0.42]{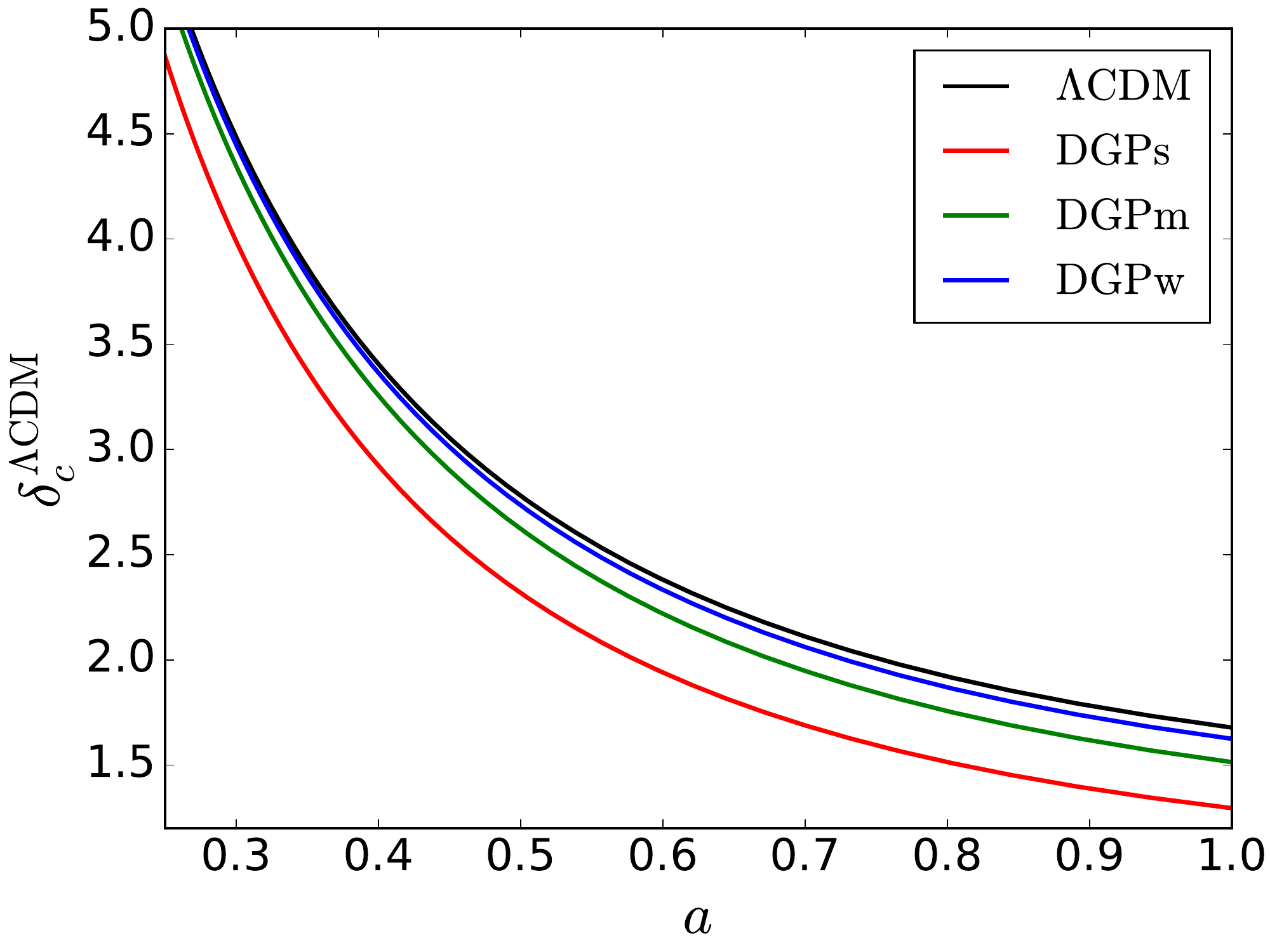}
	\caption{Time evolution of the critical density for spherical collapse $\delta_c^{\lcdm}$ {(linearly extrapolated to today assuming $\lcdm$)} for $\lcdm$ and the three DGP cases considered in this paper, as labelled.}
\label{fig:dc}
\end{figure}

In this appendix, we fit the ST mass function formulae to the simulation results. Our strategy follows closely that presented in Refs.~\cite{2013JCAP...11..056B, 2014JCAP...04..029B}, to which we refer the interested reader for more details.

The ST mass function is defined as
\bq\label{eq:mass-function}
&&\frac{{\rm d}n(M)}{{\rm d}{\rm ln}M}{\rm d}{\rm ln}M = \frac{\bar{\rho}_{m0}}{M} f(S){\rm d}S,
\eq
with $f(S)$ given by
\bq\label{eq:first-crossing-ST}
f(S) = A \sqrt{\frac{q}{2\pi}}\frac{\delta_c}{S^{3/2}}\left[1 + \left(\frac{q\delta_c^2}{S}\right)^{-p}\right]{\rm exp}\left[-q\frac{\delta_c^2}{2S}\right], \nonumber \\
\eq
where $A$ is a normalization constant fixed by the condition $\int f(S){\rm d}S = 1$, $(q,p)$ are two parameters to be fitted to the simulation results and $\delta_c \equiv \delta_c(z)$ is the critical initial overdensity for a spherical top-hat to collapse at redshift $z$, extrapolated to $z = 0$ with the $\Lambda$CDM linear growth factor. {To emphasize that $\lcdm$ is the model assumed in the linear extrapolation, we shall use the notation $\delta_c^{\lcdm}$.} The variable $S$ denotes the variance of the linear density field filtered on a comoving length scale $R$,
\bq\label{eq:variance}
S(R) \equiv \sigma^2(R) = \frac{1}{2\pi^2}\int k^2 P_{k, {\rm lin}}\tilde{W}^2\left(k, R\right){\rm d}k,
\eq
with $\tilde{W}\left(k, R\right) = 3\left({\rm sin}(kR) - kR{\rm cos}(kR)\right)/\left(kR\right)^3$. For consistency {with $\delta_c^{\lcdm}$}, $P_{k, {\rm lin}}$ in Eq.~(\ref{eq:variance}) is the initial power spectrum (which is the same for our $\lcdm$ and DGP cosmologies) also evolved to $z = 0$ with the $\Lambda$CDM linear growth factor.

To determine $\delta_c^{\lcdm}$ in the DGP model, we consider the Euler equation that governs the physical radius $r$ of the spherical overdensity \cite{2010PhRvD..81f3005S}
\bq\label{eq:sph_col}
\frac{\ddot{r}}{r} - \left(\dot{H} + H^2\right) &=& -\frac{\Psi,_{r}}{r} \nonumber \\ 
&=& -\frac{G_{\rm eff}}{G}\frac{\Psi^{\rm GR},_{r}}{r} \nonumber \\
&=&-\frac{G_{\rm eff}}{G}\frac{GM(<r)}{r^2} = - \frac{G_{\rm eff}}{G}\frac{\Omega_{m0}H_0^2a^{-3}\delta}{2}, \nonumber \\
\eq
with the effective gravitational strength $G_{\rm eff}$ given by
\bq\label{eq:sphgeff}
\frac{G_{\rm eff}(a, \delta)}{G} = 1 + \frac{2}{3\beta}\left(\frac{r}{r_V}\right)^3\left[-1 + \sqrt{1 + \left(\frac{r_V}{r}\right)^3} \right]. \nonumber \\
\eq
The second line in Eq.~(\ref{eq:sph_col}) follows from using Eq.~(\ref{eq:eomvarphi_sph2}) combined with $\Psi = \Psi^{GR} + \varphi/2$, and the third line from using the known GR result. Note that $G_{\rm eff}$ depends on both $a$ and $\delta$ (the latter entering via $r_V$, which depends on $M(<r)$ and hence on $\delta$). 

Changing the time variable to $N = {\rm ln}a$ in Eq.~(\ref{eq:sph_col}) and defining $y(t) = r(t)/\left(aR\right)$ leads to
\bq\label{eq:sph_col1}
y'' &+& \left(\frac{E'}{E} + 2\right)y' \nonumber \\
&+& \frac{G_{{\rm eff}}(a, y^{-3} - 1)}{G}\frac{\Omega_{m0}e^{-3N}}{2E^2}\left(y^{-3} - 1\right)y = 0, \nonumber \\
\eq
where $E = H/H_0$ and we have used $\delta = y^{-3} - 1$, which follows from mass conservation\footnote{Explicitly, $\bar{\rho}_ma^3R^3 = \left(1+\delta\right)\bar{\rho}_m r^3 \Rightarrow \delta = \left(aR/r\right)^3 - 1 = y^{-3} - 1$.}. The value of $\delta_c^{\lcdm}(a)$ is obtained by a trial-and-error scheme to find the initial linear density $\delta_{\rm lin, i}$ that leads to collapse ($y=0$) at scale factor $a$. The resulting value is then evolved from the initial time until $a = 1$ using the $\Lambda$CDM linear growth factor. The initial conditions are set up at $a_i = 1/300$ as $y(a_i) = 1 - \delta_{\rm lin, i}/3$ and $y'(a_i) = \delta_{\rm lin,i}/3$ (which is the matter dominated solution). The time dependence of $\delta_c^{\lcdm}$ is displayed in Fig.~\ref{fig:dc}, which shows the expected result that $\delta_c^{\lcdm}$ becomes smaller if gravity gets stronger, i.e., for the collapse to occur at the same time, then the initial overdensity must be smaller than in GR to compensate for the faster collapse. In particular, for the $\lcdm$, $\sdgp$, $\mdgp$ and $\wdgp$ models we have that $\delta_c^{\lcdm} = 2.25, 1.82, 2.09, 2.20$, respectively, at the mean redshift of the CMASS sample, $z = 0.57$ ($a \approx 0.637$).

Finally, given the simulation cumulative mass function results, $n(>M)^{\rm sims}$, we determine the best-fitting ST mass function by finding the values of $q$ and $p$ that minimize $n(>M)^{\rm sims}/n(>M, q, p)^{\rm ST} - 1$, where $n(>M, q, p)^{\rm ST}$ is the cumulative ST mass function (cf.~Eqs.~(\ref{eq:mass-function}) and (\ref{eq:first-crossing-ST})). The best-fitting values are
\bq\label{eq:bfqp}
\lcdm\ \ &::&\ \ \left(q, p\right) = \left(0.733, 0.299\right) \nonumber \\
\sdgp\ \ &::&\ \ \left(q, p\right) = \left(0.706, 0.333\right) \nonumber \\
\mdgp\ \ &::&\ \ \left(q, p\right) = \left(0.726, 0.313\right) \nonumber \\
\wdgp\ \ &::&\ \ \left(q, p\right) = \left(0.728, 0.304\right).
\eq

\section{Error reduction from cosmic variance cancellation}\label{sec:cv}

In this appendix, we derive the error reduction factor from cosmic variance cancellation $T_{\rm CV}$ used to scale the errorbars from the mock measurements of the quantity $\Delta f\sigma_8$ in Fig.~\ref{fig:fs8fit} (cf.~Eq.~(\ref{eq:errorscale})).

The estimator of the growth rate $\hat{f}$ (an overhat denotes a measured quantity to distinguish it from its expected value) can be expanded in all generality as
\bq\label{eq:fexpansion}
\hat{f} = \sum_i W_i\hat{P}(\vec{k}_i)
\eq
where $\hat{P}(\vec{k}_i) \equiv \hat{P}_i$ is the measured galaxy power spectrum in the mocks, and $W_i$ denote effective weights used in the analysis. Correspondingly,
\bq\label{eq:Dfexpansion}
\Delta\hat{f} = \sum_i W_i\left[\hat{P}_i^{\rm DGP} - \hat{P}_i^{\lcdm}\right] \equiv \sum_i W_i\Delta \hat{P}_i,
\eq
where we have assumed that the $W_i$'s are independent of the model. We are interested in estimating the variance $\Delta \hat{f}$, which is determined by the covariance of $\Delta \hat{P}$
\bq\label{eq:varF}
{\rm Var}(\Delta\hat{f}) = \sum_{ij}W_iW_j{\rm Cov}\left(\Delta \hat{P}_i, \Delta \hat{P}_j\right).
\eq
In the following, we work at linear order in cosmological perturbations so that the covariance is Gaussian. The problem is then reduced to estimating ${\rm Cov}\left(\Delta \hat{P}_i, \Delta \hat{P}_j\right)$.

\subsubsection{Estimating the covariance}\label{sec:estcov}

Using the fact that all covariances are diagonal in the Gaussian case, the covariance of $\Delta P$ can be expanded as
\bq\label{eq:covexp}
{\rm Cov}\left(\Delta \hat{P}_i, \Delta \hat{P}_j\right) &=& {\rm Cov}\left(\hat{P}_i^{\lcdm}, \hat{P}_j^{\lcdm}\right) \nonumber \\
&+& {\rm Cov}\left(\hat{P}_i^{\rm DGP}, \hat{P}_j^{\rm DGP}\right) \nonumber \\
&-& 2{\rm Cov}\left(\hat{P}_i^{\lcdm}, \hat{P}_j^{\rm DGP}\right).
\eq
The first two terms on the right-hand side are given by (we drop volume normalization factors, which are not important anyway since we shall be interested in covariance ratios)
\bq\label{eq:covauto}
{\rm Cov}\left(\hat{P}_i^{\lcdm}, \hat{P}_j^{\lcdm}\right) &=& 2\delta_{ij}\left[P_i^{\lcdm} + P_N\right]^2\nonumber \\
{\rm Cov}\left(\hat{P}_i^{\rm DGP}, \hat{P}_j^{\rm DGP}\right) &=& 2\delta_{ij}\left[P_i^{\rm DGP} + P_N\right]^2,
\eq
where $\delta_{ij}$ is the Kronecker delta and $P_N = 1/n_g^{\lcdm} \simeq 1/n_g^{\rm DGP}$ is the shot-noise galaxy power spectrum.  While $n_g^{\lcdm} = n_g^{\rm DGP}$ (cf.~Eq.~(\ref{eq:const1})), the equality of the shot noise is only approximate because of deviations from perfect Poisson shot noise.  This approximation is nevertheless sufficient for our purposes.  The cross term (last on the right-hand side of Eq.~(\ref{eq:covexp})) vanishes if the initial phases of the $\lcdm$ and ${\rm DGP}$ simulations are independent. In this case, Eq.~(\ref{eq:covexp}) becomes
\begin{widetext}
\bq
{\rm Cov}\left(\Delta \hat{P}_i, \Delta \hat{P}_j\right)_{\rm indep.\ phases} &=& 2\delta_{ij}\left[P_i^{\lcdm} + P_N\right]^2 + 2\delta_{ij}\left[P_i^{\rm DGP} + P_N\right]^2 \nonumber \\
&=& 2\delta_{ij}\left[\left(1 + \left(1 + r_i\right)^4\right)P_{i, \lcdm}^2 + 2\left(1 + \left(1 + r_i\right)^2\right)P_{i, \lcdm}P_N + 2P_N^2\right],
\eq
\end{widetext}
where we have used the fact that $P_{i, \rm DGP} = \left(1+r_i\right)^2P_{i, \lcdm}$, with
\be
1 + r_i = \sqrt{\frac{1 + \frac{2}{3} \beta^{\Lambda\rm CDM} + \frac{1}{5} (\beta^{\Lambda \rm CDM})^2}{1 + \frac{2}{3} \beta^{\rm DGP} + \frac{1}{5} (\beta^{\rm DGP})^2}}
\times\frac{1 + \beta^{\rm DGP} \mu_i^2}{1+ \beta^{\Lambda\rm CDM} \mu_i^2} ,
\label{eq:rk}
\ee
following the HOD fitting analysis of Sec.~\ref{sec:hodmodel}. When there is some positive cross-correlation (like when the simulations evolve from the same initial conditions), the total error is reduced by the cross term $-2{\rm Cov}(\hat{P}_i^{\lcdm}, \hat{P}_j^{\rm DGP})$ in Eq.~(\ref{eq:covexp}).  It is this error reduction that we wish to estimate. 
\comment{The cross term satisfies the Cauchy-Schwarz inequality
\begin{widetext}
\bq\label{eq:CSine}
{\rm Cov}\left(\hat{P}_i^{\lcdm}, \hat{P}_j^{\rm DGP}\right) \leq \sqrt{{\rm Cov}\left(\hat{P}_i^{\lcdm}, \hat{P}_i^{\lcdm}\right){\rm Cov}\left(\hat{P}_j^{\rm DGP}, \hat{P}_j^{\rm DGP}\right)}.
\eq
\end{widetext}
The best-case scenario in terms of error reduction corresponds to taking the equal sign, in which case Eq.~(\ref{eq:covexp}) becomes
\begin{widetext}
\bq
{\rm Cov}\left(\Delta \hat{P}_i, \Delta \hat{P}_j\right)_{\rm best\ case} &=& {\rm Cov}\left(\Delta \hat{P}_i, \Delta \hat{P}_j\right)_{\rm indep.\ phases} - 4\delta_{ij}\left[P_i^{\rm DGP} + P_N\right]\left[P_i^{\lcdm} + P_N\right] \nonumber \\
&=& 2\delta_{ij}\left[\left(1 + r_i\right)^2 - 1\right]^2 P_{i, \lcdm}^2.
\eq
\end{widetext}
This scenario corresponds to completely correlated galaxy density fields in the simulations (including shot noise) and is therefore too optimistic. To arrive at a more realistic estimate of the cross term,} 

Let us write this term as
\begin{widetext}
\bq\label{eq:crossfair}
{\rm Cov}\left(\hat{P}_i^{\lcdm}, \hat{P}_j^{\rm DGP}\right) &=& \langle\hat{\delta}_{i, \lcdm}^2\hat{\delta}^2_{j, {\rm DGP}}\rangle  -  \langle\hat{\delta}_{i, \lcdm}\rangle^2 \langle\hat{\delta}_{j, {\rm DGP}}\rangle^2 \nonumber \\
&=& \langle(\delta_i + \epsilon_i)^2 (\tilde{\delta}_j + \tilde{\epsilon}_j)^2\rangle - \langle(\delta_i + \epsilon_i)\rangle^2 \langle  (\tilde{\delta}_j + \tilde{\epsilon}_j)\rangle ^2 \nonumber \\
&=& 2\delta_{ij}\left[(1+r_i)P_{i, \lcdm} + \langle\epsilon\tilde{\epsilon}\rangle\right]^2
\eq
\end{widetext}
where to shorten the notation we have written the estimated density contrasts as $\hat{\delta}_{i, \lcdm} = \delta + \epsilon$ and $\hat{\delta}_{j, {\rm DGP}} = \tilde{\delta}_j + \tilde{\epsilon}_j$, with $\delta, \tilde{\delta}$ and $\epsilon, \tilde{\epsilon}$ corresponding to the deterministic and stochastic parts of the galaxy density field, respectively. We have further taken into account that $\tilde{\delta} = (1+r)\delta$, $P_{i, \lcdm}\delta_{ij} = \langle\delta_i\delta_j\rangle$, that terms like $\langle\delta\epsilon\rangle$ vanish by definition and have assumed that the noise term is diagonal $\langle\epsilon_i\tilde{\epsilon}_j\rangle = \langle\epsilon\tilde{\epsilon}\rangle \delta_{ij}$.

What is left to do is to estimate $\langle\epsilon\tilde{\epsilon}\rangle$. We can express it in terms of the halo shot noise covariance $C_N(M, M') = \langle \epsilon_M \epsilon_{M'}\rangle$ as\footnote{$\epsilon_M$ is the stochastic density constrast of the distribution of halos with mass $M$.}
\bq\label{eq:epseps}
\langle\epsilon\tilde{\epsilon}\rangle &=& \int {\rm dln}M \frac{1}{n_h}\frac{{\rm d}n_h^{\lcdm}}{{\rm dln} M}\int {\rm dln}M' \frac{1}{n_h}\frac{{\rm d}n_h^{\rm DGP}}{{\rm dln} M'} C_N(M,M') \nonumber  \\
&=& \int {\rm dln}M \frac{1}{n_h^2}\frac{{\rm d}n_h^{\lcdm}}{{\rm dln} M}\frac{{\rm d}n_h^{\rm DGP}}{{\rm dln} M}\left(\frac{{\rm d}n}{{\rm dln}M}\right)^{-1},
\eq
where we have assumed for simplicity a diagonal shot noise covariance, $C_N(M, M') = \left(\frac{{\rm d}n}{{\rm dln}M}\right)^{-1}\delta_{\rm D}\left({\rm ln}M - {\rm ln}M'\right)$.  Further, ${\rm d}n_h/{\rm dln}M$ is the mass function of halos weighted by the central halo occupation number (dashed lines in the right panel of Fig.~\ref{fig:hod}), $n_h$ is their number density, and ${\rm d}n/{\rm dln}M$ is the total halo mass function (we are not distinguishing between the $\lcdm$ and ${\rm DGP}$ halo mass functions for the moment). With the aid of Eq.~(\ref{eq:epseps}) we can define the {\it noise correlation coefficient} $\mathcal{G}_N$  as
\bq\label{eq:ncf}
\mathcal{G}_N = \frac{\langle\epsilon\tilde{\epsilon}\rangle}{\sqrt{\langle\epsilon\epsilon\rangle\langle\tilde{\epsilon}\tilde{\epsilon}\rangle}},
\eq
with 
\bq\label{eq:epseps2}
\langle\epsilon\epsilon\rangle = \int {\rm dln}M \frac{1}{n_h^2}\left(\frac{{\rm d}n_h^{\lcdm}}{{\rm dln} M}\right)^2\left(\frac{{\rm d}n}{{\rm dln}M}\right)^{-1}, \nonumber \\
\eq
and similarly for $\langle\tilde{\epsilon}\tilde{\epsilon}\rangle$. The quantity $\mathcal{G}_N$ represents a measure of the overlap in the host halo mass distribution of the $\lcdm$ and ${\rm DGP}$ mock catalogues, weighted by the shot noise $\left({{\rm d}n}/{{\rm dln}M}\right)^{-1}$. There are a number of simplifying assumptions made in this derivation, namely that (i) we neglected the random HOD sampling (which reduces $\langle\epsilon\tilde{\epsilon}\rangle$); (ii) the halo shot noise covariance is not perfectly diagonal (see \cite{hamaus/etal,schmidt:16} for a discussion) and (iii) the mass functions in $\lcdm$ and ${\rm DGP}$ are not identical. Motivated by these uncertainties, we opt for a more practical approach in which we use $\langle\epsilon\tilde{\epsilon}\rangle = \mathcal{F}_N P_N$ in Eq.~(\ref{eq:crossfair}) and treat $\mathcal{F}_N$ as a free parameter that parametrizes the noise correlation between the $\lcdm$ and ${\rm DGP}$ mock catalogues (similar to $\mathcal{G}_N$, but not exactly the same). The case $\mathcal{F}_N = 0$ corresponds to completely uncorrelated shot noise terms, which whilst being unrealistic, sets a lower bound for the improvement in error due to cosmic variance cancellation. The case $\mathcal{F}_N = 1$ corresponds to the best attainable improvement in error.

\subsubsection{Error Fisher forecast}\label{sec:fisher}

Having derived the covariance for the $\Delta P_i$, we can now perform a
Fisher forecast to determine the improvement $T_{\rm CV}$ in the error of the difference of the growth rate to $\lcdm$.  Since the covariance is diagonal, we can write it compactly as
\bq\label{eq:covcomp}
{\rm Cov}\left(\Delta \hat{P}_i, \Delta \hat{P}_j\right) = \delta_{ij}\sigma_i^2
\eq
(the shape of $\sigma_i$ is explicitly written below). The {\rm log} likelihood is given by
\bq\label{eq:loglike}
-2{\rm ln}\mathcal{L} &=& \sum_{\rm modes}\sigma_i^{-2}\left[\Delta\hat{P}_i - \langle\Delta P_i\rangle\right]^2 \nonumber \\
&=& V\int_{k_{\rm min}}^{k_{\rm max}} \frac{{\rm d}^3k}{(2\pi)^3} \frac{\left[\Delta\hat{P}(k, \mu) - \langle\Delta P(k, \mu)\rangle\right]^2}{\sigma^{2}(k, \mu)}, \nonumber \\
\eq
where in the second equality we have taken the continuum limit, $V$ is the simulation volume, $\langle\Delta P(k, \mu)\rangle$ represents the fiducial (known) difference in power spectra, $k_{\rm min} = 2\pi/V^{1/3}$ and {$k_{\rm max} = 0.050\ h/{\rm Mpc}$ is the maximum value used in the analysis (cf.~Fig.~\ref{fig:wedges}, where the minimum $s$ value modelled is $20\ {\rm Mpc}/h$).} The Fisher information for a single parameter $p$ is
\bq\label{eq:fisherinfo}
F_{pp}&\equiv&-\left\langle\frac{\partial^2}{\partial p^2}{\rm ln}\mathcal{L}\right\rangle = V\int_{k_{\rm min}}^{k_{\rm max}}\frac{{\rm d}^3k}{(2\pi)^3} \frac{\left[\frac{\partial}{\partial p}\langle\Delta P(k, \mu)\rangle\right]^2}{\sigma^2(k, \mu)} \nonumber \\
&=&\frac{V}{(2\pi)^2}\int_{k_{\rm min}}^{k_{\rm max}}k^2{\rm d}k \int_{-1}^{1}{\rm d}\mu \frac{\left[\frac{\langle\Delta P(k, \mu)\rangle}{\Delta p}\right]^2}{\sigma^2(k, \mu)},
\eq
where we have written the partial derivative w.r.t.~$p$ as a finite difference.  
A natural choice for $p$ is $\left[r_cH_0\right]^{-1}$, so that 
$\Delta p = p_{\rm DGP} - p_{\lcdm} = \left[r_cH_0\right]^{-1}$.  
Further, $\langle\Delta P(k, \mu)\rangle = \left(\left[1 + r(\mu)\right]^2 - 1\right)P_{\lcdm}(k)$.  
We use the $\wdgp$ model ($p=0.5$) to evaluate $F_{pp}$ for the following
two cases,
\begin{widetext}
\bq
\label{eq:sigmas}
\label{eq:sigmas1}\sigma^2(k,\mu)_{\rm indep.\ phases} &=& 2\delta_{ij}\left[\left(1 + \left(1 + r(\mu)\right)^4\right)P_{\lcdm}(k)^2 + 2\left(1 + \left(1 + r(\mu)\right)^2\right)P_{\lcdm}(k)P_N + 2P_N^2\right] \\
\label{eq:sigmas3}\sigma^2(k,\mu)_{\mathcal{G}_N} &=& \sigma^2(k,\mu)_{\rm indep.\ phases} - 4\delta_{ij}\left[\left(1+r_i\right)P_{i, \lcdm} + \mathcal{F}_N P_N\right]^2\,.
\eq
\end{widetext}
Finally, the factor $T_{\rm CV}$ in Eq.~(\ref{eq:errorscale}) is given by
\bq
T_{\rm CV}^{\mathcal{F}_N} &=& \sqrt{\frac{F_{pp}({\rm indep.\ phases})}{F_{pp}(\mathcal{F}_N)}}.
\eq
If $\mathcal{F}_N = 0$ in Eq.~(\ref{eq:sigmas3}) (the conservative case), then $T_{\rm CV}^{\mathcal{F}_N = 0} = 0.63$. For the $\wdgp$ model, we have $\mathcal{G}_N = 0.98$, with similar results obtained using either the $\lcdm$ or $\wdgp$ mass functions in Eq.~(\ref{eq:epseps}).   This is clearly a significant correlation, but neglects various stochastic effects as explained above.  For our best estimate, we choose $\mathcal{F}_N = 0.9$, which results in $T_{\rm CV}^{\mathcal{F}_N = 0.9} = 0.21$.

\bibliography{ndgp_rsd.bib}

\end{document}